\documentclass[lettersize,journal]{IEEEtran}
\usepackage{amsmath,amsfonts,amssymb,amsthm,bm}
\usepackage{algorithmic}
\usepackage{algorithm}
\usepackage{array}
\usepackage[caption=false,font=normalsize,labelfont=sf,textfont=sf]{subfig}
\usepackage{textcomp}
\usepackage{stfloats}
\usepackage{url}
\usepackage{verbatim}
\usepackage{graphicx}
\usepackage{cite}
\begin{document}

\title{Pilot Spoofing Attack on the Downlink of Cell-Free Massive MIMO: From the Perspective of Adversaries}

\author{Weiyang~Xu,~\IEEEmembership{Member,~IEEE}, Ruiguang~Wang\IEEEmembership{}, Yuan~Zhang\IEEEmembership{}, Hien~Quoc~Ngo,~\IEEEmembership{Senior Member,~IEEE}, and Wei~Xiang,~\IEEEmembership{Senior Member,~IEEE}
	
\thanks{W. Y. Xu, R. G. Wang and Y. Zhang are with the School of Microelectronics and Communication Engineering, Chongqing University, Chongqing, 400044, P. R. China (E-mails: \{weiyangxu, wangruiguang, 202112131047t\}@cqu.edu.cn).}
\thanks{H. Q. Ngo is with the Institute of Electronics, Communications and Information Technology (ECIT), Queen’s University Belfast, BT3 9DT, Belfast, U.K., (E-mail: hien.ngo@qub.ac.uk).}
\thanks{W. Xiang is with the School of Engineering and Mathematical Sciences, La Trobe University, Melbourne, VIC 3086, Australia (E-mail: w.xiang@latrobe.edu.au).}}

\markboth{Journal of \LaTeX\ Class Files,~Vol.~14, No.~8, August~2021}%
{Shell \MakeLowercase{\textit{et al.}}: A Sample Article Using IEEEtran.cls for IEEE Journals}


\maketitle

\begin{abstract}
The channel hardening effect is less pronounced in the cell-free massive multiple-input multiple-output (mMIMO) system compared to its cellular counterpart, making it necessary to estimate the downlink effective channel to ensure decent performance. However, the downlink training inadvertently creates an opportunity for adversarial nodes to launch pilot spoofing attacks (PSAs). First, we demonstrate that adversarial distributed access points (APs) can severely degrade the achievable downlink rate. They achieve this by estimating their channels to users in the uplink training phase and then precoding and sending the same pilot sequences as those used by legitimate APs during the downlink training phase. Then, the impact of the downlink PSA is investigated by rigorously deriving a closed-form expression of the per-user achievable downlink rate. By employing the min-max criterion to optimize the power allocation coefficients, the maximum per-user achievable rate of downlink transmission is minimized from the perspective of adversarial APs. As an alternative to the downlink PSA, adversarial APs may opt to precode random interference during the downlink data transmission in order to disrupt legitimate communications. In this scenario, the achievable downlink rate is derived, and then power optimization algorithms are also developed. We present numerical results to showcase the detrimental impact of the downlink PSA and compare the effects of these two types of attacks.
\end{abstract}


\section{Introduction}
\IEEEPARstart{C}{ell-free} massive multiple-input multiple-output (mMIMO) systems are a distributed network consisting of a large number of randomly located access points (APs) \cite{Ngo}. Compared to its cellular counterpart, the cell-free mMIMO system provides ubiquitous communications with high spectral efficiency thanks to its additional macro-diversity and greater ability of interference suppression. Moreover, cell-free mMIMO is scalable in the sense that the signal processing and fronthaul signaling tasks remain feasible when the number of users in the network increases \cite{Elhoushy}. Hence, cell-free mMIMO is regarded as a promising physical layer technique for Beyond 5G (B5G) and towards Sixth-Generation (6G) networks.

On the other hand, due to their broadcast nature, wireless communications are vulnerable to adversarial attacks. Traditional methods for security are to implement cryptographic encryption in the application layer. However, this approach is potentially insecure as it is based on the assumption of computational complexity \cite{mukherjee}. As an alternative, physical layer security has become one of effective means to realize secure communications \cite{trappe}. Rather than resorting to high-level cryptographic methods, physical layer security techniques employ information-theoretic security and signal processing techniques. Generally, passive and active attacks are the two major concerns in this context. In particular, cell-free mMIMO can dramatically boost security against passive eavesdropping thanks to its inherited capability from cellular mMIMO to concentrate the transmission energy in the direction of legitimate users \cite{kapetanovic}. However, when an eavesdropper launches active attacks, the secrecy rate will be dramatically reduced. For example, the channel state information (CSI), which is crucial for exploiting the benefits of cell-free mMIMO, is generally estimated by sending pilots ahead of payload data transmission \cite{hoang}. Nevertheless, this training phase creates an opportunity for adversarial nodes to launch attacks. By sending the same pilots as legitimate users do, the channel estimates can be contaminated, resulting in severe information leakage on the downlink transmission \cite{xwy}. Such a mechanism, referred to as pilot spoofing attacks (PSAs), was first documented in \cite{zhouxy} and has received a great deal of attention since then.

\subsection{Related Work}
Cell-free mMIMO, like its cellular counterpart, is incredibly susceptible to PSAs. As a result, significant efforts have gone into developing its detection methods and countermeasures, and useful algorithms have been developed. In \cite{hoang}, an energy-based method to detect the presence of PSAs in cell-free mMIMO was proposed, and then path-following algorithms were developed to solve an optimization problem aiming at maximizing the achievable rate of legitimate users. More recently, the authors in \cite{10105654} presented the first performance analysis of physical layer downlink secure transmission in a scalable cell-free mMIMO system, where stochastic geometry was used to model the node locations. The secrecy energy efficiency optimization problem was studied in multi-user multi-eavesdropper cell-free mMIMO networks, where a confidential and energy-efficient design for transmit power allocation was developed \cite{10016749}. For the downlink of cell-free mMIMO, reference \cite{10114982} investigated the threat of passive eavesdropping on downlink cell-free mMIMO systems. Artificial noise was employed to jam the eavesdropper's signal under the effect of imperfect channel estimation. The angle-domain filtering method was developed in \cite{Qiu} to reduce the eavesdropping and interference from illegal users, thereby improving the secure transmission. 

More recently, the impact of radio frequency (RF) impairments on the ergodic secrecy rate of cell-free mMIMO systems was evaluated, and compensation algorithms for these nonidealities were proposed in \cite{x.zhang}. While the authors of \cite{s.elhoushy} analyzed the potential of the reconfigurable intelligent surface (RIS) in boosting the secrecy capacity of cell-free mMIMO systems under PSAs, where the power coefficients at APs and RIS phase shifts were jointly optimized. Addressing the problem of information leakage in user-centric cell-free mMIMO system, the precoding was optimized via formulating a secrecy rate maximization problem under the minimum rate requirements of users and the power constraints of APs \cite{gao}. Besides, it is worth noting that due to the similarities between cellular and cell-free mMIMO systems, some algorithms originally designed for cellular mMIMO are still applicable to cell-free MIMO systems \cite{Tugnait}.

\subsection{Motivation and Contributions}
We draw attention to the fact that current research focuses on PSAs in uplink training—that is, when uplink pilots are being transmitted. This is because the data detection on the downlink of cellular mMIMO relies on statistical CSI, so the downlink training phase is often unnecessary \cite{marzetta_larsson_yang_ngo_2016}. This is manifested by the phenomenon called {\it channel hardening}, which is observed at the receiver when a signal is transmitted by a large number of antennas \cite{z.chen}. Since the channel hardening effect is not as strong as it is in cellular mMIMO scenarios, this approach is not favored in cell-free mMIMO networks. In order to considerably increase the achievable rate for cell-free mMIMO systems, the concept of downlink training was introduced in \cite{inter}.

The downlink training, however, brings about a fresh issue. Despite its advantages, it unintentionally gives adversarial nodes a chance to launch PSAs. Our work is primarily driven by the need to comprehend how the PSA affects the achievable downlink rate during the downlink training phase. To the best of our knowledge, this work is the first to examine downlink PSAs in cell-free mMIMO networks. The main contributions are summarized as follows.
\begin{itemize}
	\item Modeling and analysis are carried out to determine how the downlink PSA will affect the cell-free mMIMO system. With regards to the achievable downlink rate in the presence of PSAs, a closed-form expression is developed. A performance analysis examining how the achievable rate varies with the key system parameters is conducted.	
	\item To minimize the maximum per-user achievable rate of downlink transmission, the power allocation coefficients of adversarial APs are optimized by using the min-max criterion. In particular, the downlink per-user achievable rate provided by the optimized coefficients is compared with that of equal power allocation.
	\item Furthermore, in lieu of launching downlink PSAs, we propose to let adversarial APs send precoded random interference during the downlink data transmission phase to disrupt legitimate communications. Similarly, the corresponding min-max power allocation problem is investigated. Results show that with a given transmit power budget, attacking the downlink data transmission phase is more effective in terms of reducing the achievable rate.
\end{itemize}

The remainder of this paper is organized as follows. The considered system model is illustrated in Section \ref{II}. The description of downlink PSA is detailed in Section \ref{III}. Section \ref{IIII} presents the achievable downlink rate analysis and optimal power allocation from the perspective of adversarial APs. The achievable rate analysis and power allocation with respect to attacking the downlink data transmission phase are presented in Section \ref{IIIII}. Numerical simulations are conducted to validate our analysis in Section \ref{IIIIII}. Finally, concluding remarks are made in Section \ref{conclusion}.

$Notation$: $\mathbb{C}^{n \times m}$ indicates a complex matrix of dimension $n \times m$. Bold variables represent matrices and vectors. Random variable $x \sim {\cal CN}\left(\mu,\sigma^2\right)$ denotes a complex Gaussian distribution with mean $\mu$ and variance $\sigma^2$. ${\left(\cdot\right)^T}$, ${\left(\cdot\right)^H}$, ${\left(\cdot\right)^*}$, and $\left \| \cdot \right \|^2_2$ refer to the transpose, conjugate transpose, complex conjugate, and ${\cal L}_2$ norm operators, respectively. $[\mathbf{A}]_{mn}$ indicates the element of the $m$-th row and $n$-th column of matrix $\mathbf{A}$. Finally, ${\mathbb E}[\cdot]$, ${\text{var}}[\cdot]$, and ${\text{cov}}[\cdot]$ are taken to mean the expectation, variance, and covariance operators, respectively. 

%$Notation$: $\mathbb{C}^{n \times m}$ indicates a complex matrix of size $n \times m$. Bold variables represent matrices and vectors. For a random variable $x$, $x \sim {\cal CN}(\mu,\sigma^2)$ and $x \sim {\cal N}(\mu,\sigma^2)$ indicate complex and real Gaussian distributions with mean $\mu$ and variance $\sigma^2$, respectively. ${\left(\cdot\right)^T}$, ${\left(\cdot\right)^{\rm H}}$, ${\left(\cdot\right)^*}$ and $\left \| \cdot \right \|^2_2$ denote the transpose, conjugate transpose, complex conjugate and ${\cal L}_2$ norm operators. $\Re\{\cdot\}$ and $\Im\{\cdot\}$ refer to the real and imaginary parts of complex numbers.  $\text{erf}(\cdot)$ and $\text{erfc}(\cdot)$ separately represents the error and complementary error functions. Finally, ${\mathbb E}[\cdot]$, ${\text{var}}[\cdot]$ and ${\text{cov}}[\cdot]$ indicate the expectation, variance, and covariance operators, respectively.

\section{System Model Description}\label{II}
We consider a cell-free mMIMO network with $M$ APs and $K$ users. All APs and users are equipped with a single antenna and randomly located in a large area. Besides, the APs are connected to a central processing unit (CPU) via a backhaul network. It is assumed that $M$ APs simultaneously serve $K$ users using the same time-frequency resources. In particular, the channel between the $m$-th AP and the $k$-th user is denoted by
\begin{equation}
	g_{mk} = h_{mk}\sqrt{\beta_{mk}},
\end{equation}
where $h_{mk}\sim {\cal CN}\left(0,1\right)$ is the small-scale fading coefficient, and $\beta_{mk}$ indicates the large-scale fading coefficient, which models the path-loss and shadowing effects. Since $\beta_{mk}$ fluctuates slowly and can be accurately estimated and tracked, it is assumed that the APs and users have perfect knowledge of these coefficients. In addition, all nodes are supposed to be perfectly synchronized and operate in the time-duplex division (TDD) mode. Each TDD coherence interval is divided into four phases: uplink training, uplink data transmission, downlink training, and downlink data transmission.

\subsection{Uplink Training}
First, we provide a quick summary of the uplink training. Denoted by $\bm{\varphi}_k \in \mathbb{C}^{\tau_{\rm u} \times 1}$, $k = 1,\cdots,K$, the uplink pilot sequence assigned to the $k$-th user, with $\tau_{\rm u}$ being the pilot length. It is assumed that the pilot sequences assigned to different users are mutually orthonormal, i.e., $\bm{\varphi}_i^{\rm H} \bm{\varphi}_j = \delta_{ij}$, where $\delta_{ij}$ denotes the Kronecker delta.

After channel propagation, the received $\tau_{\rm u} \times 1$ pilot vector at the $m$-th AP is given by
\begin{equation}\label{ul_rec_pilot}
	\bm{y}_{{\rm up},m}= \sqrt{\tau_{\rm u} \rho_{\rm up}} \sum\limits_{k=1}^K g_{mk} \bm{\varphi}_k + \bm{w}_{{\rm up},m},
\end{equation}
where the subscript ``up'' denotes uplink pilots, $\rho_{\rm up}$ is the normalized transmit signal-to-noise ratio (SNR) of uplink pilots, and $\bm{w}_{{\rm up},m}$ is the additive noise vector with its elements obeying a distribution of $\mathcal{C} \mathcal{N}(0, 1)$. The $m$-th AP then projects $\bm{y}_{{\rm up},m}$ onto $\bm{\varphi}_k^{\rm H}$ and estimates the channel coefficient using the minimum mean square error (MMSE) method. The channel estimate of $g_{mk}$ is given by
\begin{equation}\label{channel-estimate-bob}
	\hat{g}_{mk} = \frac{\sqrt{\tau_{\rm u} \rho_{\rm up}} \beta_{mk}}{1+\tau_{\rm u} \rho_{\rm up} \beta_{mk}}\bm{\varphi}_k^{\rm H} \bm{y}_{{\rm up},m}.
\end{equation}
Denoting by ${\tilde{g}_{mk}} \triangleq g_{mk} - {\hat{g}_{mk}}$ the channel estimation error, we have
\begin{equation}
	\begin{aligned}
	{\hat{g}_{mk}} &\sim \mathcal{C} \mathcal{N}\left(0,\gamma_{mk}\right),\\
	 {\tilde{g}_{mk}} &\sim \mathcal{C} \mathcal{N}\left(0,\beta_{mk} - \gamma_{mk}\right),
	 \end{aligned}
\end{equation}
where $\gamma_{mk} = \frac{\rho_{\rm up} \tau_{\rm u} \beta_{mk}^2}{1+\rho_{\rm up}\tau_{\rm u} \beta_{mk}}$. Attributed to the property of MMSE estimation, ${\tilde{g}_{mk}}$ and ${\hat{g}_{mk}}$ are mutually uncorrelated.

\subsection{Downlink Training with Beamforming}
During this phase, the downlink pilot sequences are beamformed to users using conjugate beamforming. Similarly, let $\bm{\phi}_k \in \mathbb{C}^{\tau_{\rm d} \times 1}$ be the downlink pilot sequence used by the $k$-th user, where $\bm{\phi}_i^{\rm H} \bm{\phi}_j = \delta_{ij}$. Hence, the $\tau_{\rm d} \times 1$ downlink pilot vector to be transmitted by the $m$-th AP is given by \cite{8334259}
\begin{equation}
	\bm{x}_{{\rm dp},m} = \sqrt{\tau_{\rm d} \rho_{\rm dp}} \sum\limits_{k=1}^K \sqrt{\eta_{mk}}\hat{g}^*_{mk}\bm{\phi}_k,
\end{equation}
where the subscript ``dp'' denotes downlink pilots, $\rho_{\rm dp}$ is the normalized transmit SNR of the downlink pilots, and $\eta_{mk}$ is the power coefficient used by the $m$-th AP for transmission to the $k$-th user. Therefore, the $\tau_{\rm d} \times 1$ downlink pilot vector received by the $k$-th user is
\begin{equation}\label{dl-rec-pilots}
	\bm{y}_{{\rm dp},k} = \sqrt{\tau_{\rm d} \rho_{\rm dp}} \sum\limits_{k'=1}^K a_{kk'} \bm{\phi}_{k'} + \bm{w}_{{\rm dp},k},
\end{equation}
where $\bm{w}_{{\rm dp},k}$ is the noise vector and its element has the same distribution as that of $\bm{w}_{{\rm up},m}$, and $a_{kk'} = \sum\nolimits_{m=1}^M \sqrt{\eta_{mk'}} g_{mk} \hat{g}^*_{mk'}$. In particular, $a_{kk}$ describes the {\it effective downlink channel} and can be estimated by first projecting $\bm{y}_{{\rm dp},k}$ onto pilot sequence $\bm{\phi}_k^{\rm H}$ to obtain ${y}_{{{\rm dp},k}} = \bm{\phi}_k^{\rm H} \bm{y}_{{\rm dp},k}$, and then applying the MMSE channel estimation method. Therefore, the estimation results of $a_{kk}$ is obtained as follows \cite{inter}
\begin{equation}\label{dl-channel-estimate}
	\hat{a}_{kk} = \mathbb{E}\left\{a_{kk}\right\} + \frac{{\rm cov}\left\{a_{kk},{y}_{{{\rm dp},k}}\right\}}{{\rm cov}\left\{{y}_{{{\rm dp},k}},{y}_{{{\rm dp},k}}\right\}}\left({y}_{{{\rm dp},k}}-\mathbb{E}\left\{{y}_{{{\rm dp},k}}\right\}\right),
\end{equation}
where
\begin{equation}\label{constants}
	\begin{aligned}
		\mathbb{E} \left\{a_{kk}\right\} & = \sum\limits_{m=1}^M \sqrt{\eta_{mk}} \gamma_{mk},\\
		\mathbb{E} \left\{{y}_{{{\rm dp},k}}\right\} & = \sqrt{\tau_{\mathrm{d}}\rho_{\mathrm{dp}}} \sum\limits_{m=1}^M \sqrt{\eta_{mk}}\gamma_{mk},\\
		\mathrm{cov} \left\{a_{kk},{y}_{{{\rm dp},k}}\right\} & = \sqrt{\tau_{\mathrm{d}}\rho_{\mathrm{dp}}} \sum\limits_{m=1}^M \eta_{mk}\gamma_{mk}\beta_{mk},\\
		\mathrm{cov} \left\{{y}_{{{\rm dp},k}},{y}_{{{\rm dp},k}}\right\} & = 1 + \tau_{\mathrm{d}} \rho_{\mathrm{dp}} \eta_{mk} \gamma_{mk}.
	\end{aligned}
\end{equation}
The channel estimation error is given by $\tilde{a}_{kk} = a_{kk} - \hat{a}_{kk}$, which is uncorrelated with the corresponding channel estimate, just as in the case of uplink training. Despite an increase in the per-user achievable rate, we emphasize that the downlink training phase poses a possible threat to legitimate transmission, as will be demonstrated below.

\section{PSA in the Downlink Training Phase}\label{III}
%In this section, we will demonstrate that the downlink PSA can alter the effective channel estimate result and thus reduce the achievable rate of legitimate users. Afterwards, a closed-form expression of achievable downlink rate in the presence of PSA is derived.

\begin{figure}
	\centering
	\includegraphics[width=80mm]{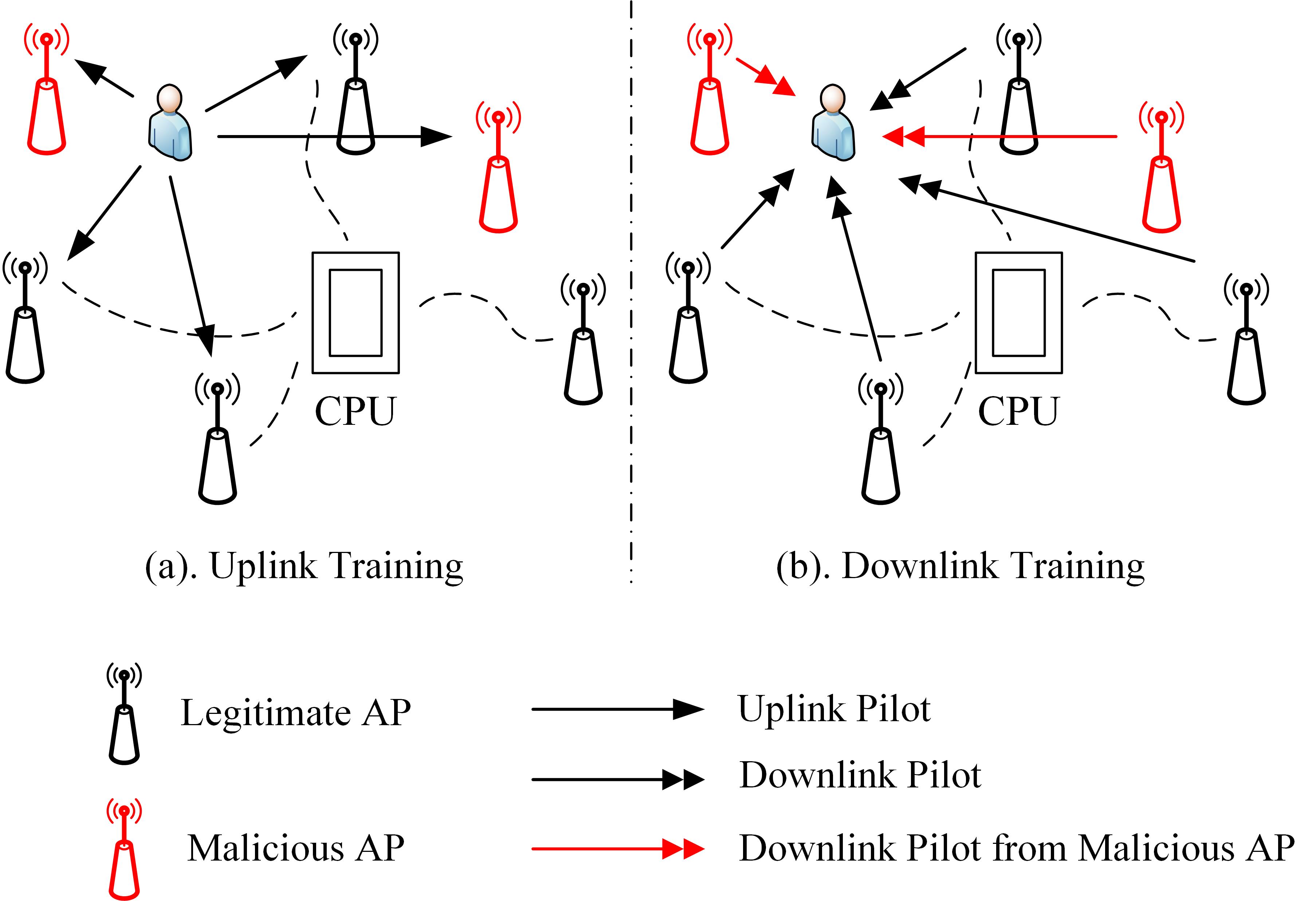}
	\caption{Illustration of the downlink PSA, (a) Malicious APs carry out channel estimation during the uplink training phase; (b) Malicious APs send the beamformed downlink pilot sequences to users.}
	\label{fig1}
\end{figure}

Suppose there are $N$ adversarial APs distributed in the same region as legitimate APs, as depicted in Fig. \ref{fig1}. In particular, the channel between the $n$-th adversarial AP and the $k$-th user is modeled as
\begin{equation}
	f_{nk} = q_{nk}\sqrt{\theta_{nk}},
\end{equation}
where $q_{nk} \sim \mathcal{CN}(0,1)$ is the small-scale fading factor, whilst $\theta_{nk}$ is the large-scale fading coefficient and known \textit{a priori}. In principle, adversarial APs should estimate the channel in the uplink training phase and utilize this information to precode the subsequent downlink pilot sequences in the downlink training phase in order to launch downlink PSAs. In the ensuing sections, we will go through these two steps in further detail.

As the first step, the adversarial APs employ uplink pilot sequences to assess their channel toward users because they are publicly available. Hence, the received $\tau_{\rm u} \times 1$ pilot vector at the $n$-th adversarial AP is given by
\begin{equation}\label{ul_rec_pilot}
	\bm{y}_{{\rm up},n}= \sqrt{\tau_{\rm u} \rho_{\rm up}} \sum\limits_{k=1}^K f_{nk} \bm{\varphi}_k + \bm{w}_{{\rm up},n},
\end{equation}
Similar to \eqref{channel-estimate-bob}, the $n$-th adversarial AP calculates the channel coefficient of the $k$-th user using the MMSE criterion, i.e.,
\begin{equation}\label{channel-estimate-eve}
	\hat{f}_{nk} = \frac{\sqrt{\tau_{\rm u} \rho_{\rm up}} \theta_{nk}}{1+\tau_{\rm u} \rho_{\rm up} \theta_{nk}}\bm{\varphi}_k^{\rm H} \bm{y}_{{\rm up},n}.
\end{equation}
Similar to that in legitimate communications, the uplink channel estimation error, defined as ${\tilde{f}_{nk}} \triangleq f_{nk} - {\hat{f}_{nk}}$, is uncorrelated with ${\hat{f}_{nk}}$. Moreover, it is derived that ${\hat{f}_{nk}} \sim \mathcal{C} \mathcal{N}\left(0,\kappa_{nk}\right)$ and ${\tilde{f}_{nk}} \sim \mathcal{C} \mathcal{N}\left(0,\theta_{nk} - \kappa_{nk}\right)$, where $\kappa_{nk} = \frac{\rho_{\rm up} \tau_{\rm u} \theta_{nk}^2}{1+\rho_{\rm up}\tau_{\rm u} \theta_{nk}}$.

In the second step, the adversarial APs exploit conjugate beamforming to precode and transmit downlink pilot sequences to users. It should be noted that using beamforming systems other than those used by legitimate APs could significantly complicate our analysis, which is not helpful for obtaining an in-depth understanding of the downlink PSA. As a result, conjugate beamforming is used and the $\tau_{\rm d} \times 1$ received downlink pilot vector of the $n$-th adversarial AP is given by
\begin{equation}\label{eve_precoding}
	\bm{x}_{{\rm dp},n} = \sqrt{\tau_{\rm d} \mu_{\rm dp}} \sum\limits_{k=1}^K \sqrt{\zeta_{nk}}\hat{f}^*_{nk}\bm{\phi}_k,
\end{equation}
where $\mu_{\rm dp}$ is the normalized transmit SNR of the downlink pilot of adversarial APs, and $\zeta_{nk}$ denotes the power allocation factor of the $n$-th adversarial AP for transmitting $\bm{\phi}_k$. Since both the legitimate and adversarial APs send beamformed pilot sequences simultaneously and synchronously, then \eqref{dl-rec-pilots} is rewritten as
\begin{equation}\label{dl-rec-pilots-eve}
	\bar{\bm{ y}}_{{\rm dp},k} = \sqrt{\tau_{\rm d} \rho_{\rm dp}} \sum\limits_{k'=1}^K a_{kk'} \bm{\phi}_{k'} +\sqrt{\tau_{\rm d} \mu_{\rm dp}} \sum\limits_{k'=1}^K b_{kk'} \bm{\phi}_{k'} + \bm{w}_{{\rm dp},k},
\end{equation}
where 
\begin{equation*}
	b_{kk'} = \sum\limits_{n=1}^N \sqrt{\zeta_{nk'}} f_{nk} \hat{f}^*_{nk'},\quad k'=1,\dots,K,
\end{equation*}
and we use $\bar{\bm{ y}}_{{\rm dp},k}$ to denote $\bm{ y}_{{\rm dp},k}$ in the presence of PSAs. It is important to note that the second component in \eqref{dl-rec-pilots-eve} represents the interference from the adversarial APs.

The $k$-th user estimates the downlink effective channel using \eqref{dl-channel-estimate}, because it is unaware of the existence of the downlink PSA. Detection of the downlink PSA is beyond the scope of this paper. Because the expectations and covariances in \eqref{constants} depend on known statistics, they can be calculated and stored in advance to facilitate channel estimation. Therefore, the received signal is the only source of uncertainty in \eqref{dl-channel-estimate}. In the presence of the downlink PSA, ${y}_{{{\rm dp},k}}$ can be rewritten as
\begin{equation}
	\begin{aligned}
		\bar{y}_{{{\rm dp},k}} &= \bm{\phi}_k^{\rm H} \bar{\bm{ y}}_{{\rm dp},k} \\
				&= \sqrt{\tau_{\rm d} \rho_{\rm dp}} a_{kk} + \sqrt{\tau_{\rm d} \mu_{\rm dp}} b_{kk} + {n}_{{\rm dp},k},
	\end{aligned}	
\end{equation}
where ${n}_{{\rm dp},k} = \bm{\phi}_k^{\rm H} \bm{w}_{{\rm dp},k}$, and we use $\bar{{ y}}_{{\rm dp},k}$ to denote ${ y}_{{\rm dp},k}$ in the presence of PSAs. By replacing ${y}_{{{\rm dp},k}}$ with $\bar{y}_{{{\rm dp},k}}$ in \eqref{dl-channel-estimate}, one can obtain the estimation result as
\begin{equation}\label{channel-estimate-eve-1}
	\hat{\bar{a}}_{kk} = \mathbb{E}\left\{a_{kk}\right\} + \frac{{\rm Cov}\left\{a_{kk},{y}_{{{\rm dp},k}}\right\}}{{\rm Cov}\left\{{y}_{{{\rm dp},k}},{y}_{{{\rm dp},k}}\right\}}\left(\bar{y}_{{{\rm dp},k}}-\mathbb{E}\left\{{y}_{{{\rm dp},k}}\right\}\right),
\end{equation}
where $\hat{\bar{a}}_{kk}$ is the estimate of the effective downlink channel in the presence of the PSA. Comparing \eqref{dl-channel-estimate} with \eqref{channel-estimate-eve-1} leads to the discovery that except for $\bar{y}_{{\rm dp},k}$, the other parameters remain unaltered because users are unaware that the received signals contain pilots sent by adversarial APs. However, this seemingly insignificant difference can have a significant impact on system performance.

{\it Remark 1:} Due to the existence of $\sqrt{\tau_{\rm d} \mu_{\rm dp}} b_{kk}$ in $\bar{y}_{{\rm dp},k}$, the channel estimation result includes not only the desired channel $a_{kk}$, but also the channel with respect to adversarial APs. We point out that simply boosting the transmit power of legal APs would not eliminate the interference. If users perform data decoding using the contaminated channel estimate, there could be a considerable loss in the achievable downlink rate. Additionally, adversarial APs may act in collusion to optimize the power allocation factor $\zeta_{nk}$, and thus further reduce the downlink rate. Hence, the downlink PSA poses a severe threat to the security of cell-free mMIMO systems.

{\it Remark 2:} In addition to the aforementioned tactic, adversarial APs have a number of potential choices to impact legitimate communications. For example, adversarial APs can decide to just interfere with a subset of users rather than attacking all of them. This is achieved by setting $\zeta_{nk} = 0$ in \eqref{eve_precoding} if user $k$ is not targeted. For users who are being targeted, the attack may result in a significant rate loss and even outage. Besides, adversarial APs can attack not only the downlink training phase but also the downlink data transmission phase. By precoding random interference signals and sending them to users, the signal-to-interference-plus-noise ratio (SINR) of legitimate communications would be further degraded, as will be elaborated on in more depth later.

\section{Downlink Achievable Rate Analysis and Power Allocation}\label{IIII}
\subsection{Downlink Achievable Rate Analysis}
In this section, we derive the per-user achievable downlink rate in the presence of downlink PSAs. During the downlink data transmission phase, each legitimate AP employs its estimated CSI to precode the payload data symbols. On the contrary, adversarial APs remain silent in this interval. With conjugate beamforming, the signal transmitted by the $m$-th AP to all users is
\begin{equation}\label{pcoding-data}
	{x_{{\rm d},m}} = \sqrt {{\rho _{\rm d}}} \sum\limits_{k = 1}^K {\sqrt {{\eta _{mk}}} {{\hat g}^*}_{mk}{s_k}},
\end{equation}
where $\rho _{\rm d}$ is the normalized transmit SNR, $s_k$ denotes the transmit symbol for the $k$-th user and we assume that $\mathbb{E}\small\{\left|s_k\right|^2\small\} = 1$. After channel propagation, the $k$-th user receives a linear combination of signals transmitted by all legitimate APs, i.e.,
\begin{equation}\label{dlsignal}
	\begin{aligned}
		{r_{{\rm d},k}} &= \sum\limits_{m = 1}^M {{g_{mk}}{x_{{\rm d},m}}}  + {w_{{\rm d},k}}\\
		&= \sqrt {{\rho _{\rm d}}} \sum\limits_{m = 1}^M {\sum\limits_{{k'} =1}^K {\sqrt {{\eta _{m{k'}}}} {g_{mk}}} {{\hat g}^*}_{m{k'}}{s_{{k'}}}}  + {w_{{\rm d},k}}\\
		&= \sqrt {{\rho _{\rm d}}} {a_{kk}}{s_k} + \sqrt {{\rho _{\rm d}}} \sum\limits_{{k'} \ne k}^K {{a_{kk'}}{s_{{k'}}}}  + {w_{{\rm d},k}}.
	\end{aligned}
\end{equation}
In what follows, the mutual information between the received signal $r_{{\rm d},k}$ and the transmitted symbol $s_k$ is exploited to derive the per-user achievable downlink rate.

Denoted by $\tilde {\bar{a}}_{kk}$ the estimation error of the effective channel in the presence of downlink PSAs. Then, $a_{kk}$ can be written as
\begin{equation}\label{channel-estimate}
	{a_{kk}} = {\tilde {\bar{a}}_{kk}} + {\hat {\bar{a}}_{kk}}.
\end{equation}
As the linear MMSE method is adopted, the estimated channel $ {\hat {\bar{a}}_{kk}}$ and estimation error ${\tilde {\bar{a}}_{kk}}$ are uncorrelated. However, they are not independent because they are not Gaussian distributed. To derive the achievable downlink rate, the signal seen by the $k$-th user in \eqref{dlsignal} is first rewritten as
\begin{equation}
	{r_{{\rm d},k}} = \sqrt {{\rho _{\rm d}}} {a_{kk}}{s_k} + {\tilde w_{{\rm d},k}},
\end{equation}
where ${\tilde w_{{\rm d},k}} = \sqrt {{\rho _{\rm d}}} \sum\nolimits_{m = 1}^M \sum\nolimits_{k'\ne k}^{K} \sqrt{\eta_{mk'}}g_{mk}{\hat{g}^*}_{mk'}s_{k'} + w_{{\rm d},k}$ denotes the effective noise. Since $s_k$ is of zero mean and unit variance, it follows that \cite{inter}
\begin{equation}\label{zero_corre}
	\begin{aligned}		
		\mathbb{E}\left\{ {{{\tilde w}_{{\rm d},k}}\left| {{{\hat {\bar{a}}}_{kk}}} \right.} \right\} &= \mathbb{E}\left\{ {s_k^*{{\tilde w}_{{\rm d},k}}\left| {{{\hat {\bar{a}}}_{kk}}} \right.} \right\}\\
		&=\mathbb{E}\left\{ a_{kk}^*{s_k^*{{\tilde w}_{{\rm d},k}}\left| {{{\hat {\bar{a}}}_{kk}}} \right.} \right\}=0.
	\end{aligned}
\end{equation}
Then according to \cite{m.medard}, the achievable downlink rate of user $k$ is computed by
\begin{equation}\label{r_k}
	{R_k} = \mathbb{E}\left\{\log_2\left(1 + \mathrm{SINR}_{k}\right)\right\},
\end{equation}
where
\begin{equation*}
	\mathrm{SINR}_{k} = \frac{{{{\left| {\mathbb{E}\left\{ {{a_{kk}}| {{{\hat {\bar{a}}}_{kk}}}} \right\}} \right|}^2}}}{ \sum\limits_{k'=1}^{K}\mathbb{E}\left\{\left|a_{kk'}\right|^2|\hat{{\bar{a}}}_{kk}\right\} - {{{\left| {\mathbb{E}\left\{ {{a_{kk}}| {{{\hat {\bar{a}}}_{kk}}}} \right\}} \right|}^2}} + \frac{1}{\rho_{\rm d}}}
\end{equation*}
represents the received SINR of the $k$-th user.

\newcounter{mytempeqncnt}
\begin{figure*}[!t]	
	\setcounter{mytempeqncnt}{\value{equation}}	
	\setcounter{equation}{25}
	\begin{equation}\label{fr}
		{R_k} \triangleq \log_2\left(1+\mathrm{SINR}''_{k}\right)\approx {\log _2}\left(1 + \frac{{C\sum\limits_{n = 1}^N {\sqrt {{\zeta _{nk}}} } {\kappa _{nk}} + \underbrace{{D\sum\limits_{n = 1}^N {{\zeta _{nk}}} {\theta _{nk}}{\kappa _{nk}} + D{{\left(\sum\limits_{n = 1}^N {\sqrt {{\zeta _{nk}}} } {\kappa _{nk}}\right)}^2}}}_{\mathcal{D}} + A}}{\underbrace{{D\sum\limits_{n = 1}^N {{\zeta _{nk}}} {\theta _{nk}}{\kappa _{nk}} + D{{\left(\sum\limits_{n = 1}^N {\sqrt {{\zeta _{nk}}} } {\kappa _{nk}}\right)}^2}}}_{\mathcal{D}} + B}\right),
	\end{equation}
	\setcounter{equation}{\value{mytempeqncnt}}	
	\hrulefill	
	\vspace*{4pt}	
\end{figure*}
The derivation of \eqref{r_k} is quite lengthy due to the complex form of ${\hat {\bar{a}}_{kk}}$ shown in \eqref{channel-estimate-eve-1}. As a result, we use approximations to simplify the derivation process. In particular, we note that $a_{kk'}$ is the sum of independent distributed random variables. Hence, it can be approximated as Gaussian variables as $M \to \infty$ according to the Cramér central limit theorem, i.e.,
\begin{equation}\label{gs-appr}
	\begin{aligned}
		{a_{kk'}} &\mathop  \to \limits^\mathrm{d} \mathcal{CN}\left( {\bm{\varphi}_k^{\rm H}\bm{\varphi}_{k'}\sum\limits_{m = 1}^M {\sqrt {{\eta _{mk'}}} {\gamma _{mk'}}\frac{{{\beta _{mk}}}}{{{\beta _{mk'}}}}} ,{\varsigma _{kk'}}} \right),~ k \ne k'\\
		{a_{kk}} & \mathop  \to \limits^\mathrm{d} \mathcal{CN}\left( {\sum\limits_{m = 1}^M {\sqrt {{\eta _{mk}}} {\gamma _{mk}}} ,{\varsigma _{kk}}} \right),
	\end{aligned}
\end{equation}
where ${\varsigma _{kk'}} = \sum\nolimits_{m = 1}^M {{\eta _{mk'}}} {\beta _{mk}}{\gamma _{mk'}}$, and $\mathop  \to \limits^\mathrm{d} $ denotes the convergence in distribution. A tight match between the empirical and Gaussian distributions was verified even for small $M$ in \cite{inter}, supporting the validity of approximations in \eqref{gs-appr}. Additionally, the imaginary part of $a_{kk}$ is significantly smaller than its real counterpart and thus can be disregarded, that is, ${a_{kk}} \mathop  \to \limits^\mathrm{d} \mathcal{N}\left( {\sum\nolimits_{m = 1}^M {\sqrt {{\eta _{mk}}} {\gamma _{mk}}} ,{\varsigma _{kk}}} \right)$.
% Since the pilot sequences are mutually orthogonal, the distribution of $a_{kk'}$ is simplified to ${a_{kk'}} \mathop  \to \limits^d \mathcal{CN} \left( {0,{\varsigma _{kk'}}} \right)$. 

Given that $a_{kk}$ follows a Gaussian distribution, we arrive that ${\hat {\bar{a}}_{kk}}$ and $\tilde{\bar{a}}_{kk}$ are mutually independent. The same method can be used to demonstrate that any linear combination of $a_{kk}$ and $a_{kk'}$ is asymptotically Gaussian-distributed (for large values of $M$). Thus, $a_{kk}$ and $a_{kk'}$ are asymptotically joint Gaussian distributions. Therefore, the achievable downlink rate in \eqref{r_k} can be approximated to
\begin{equation}\label{r_k1}
	R_k \approx \mathbb{E}\left\{\log_2\left(1 + \mathrm{SINR}'_{k}\right)\right\}
\end{equation}
where
\begin{equation*}
	\mathrm{SINR}'_{k} = \frac{{{\rho _{\rm d}}{{\left| {{{\hat{\bar{a}}}_{kk}}} \right|}^2}}}{{{\rho _{\rm d}}\mathbb{E}\small\{ {{{\left| {{{\tilde{\bar{a}}}_{kk}}} \right|}^2}} \small\} + \rho_{\rm d}\sum\nolimits_{k'\ne k}^{K}\mathbb{E}\small\{\left|a_{kk'}\right|^2|\hat{{\bar{a}}}_{kk}\small\} + 1 }}.
\end{equation*}
To obtain a closed-form expression of $R_k$, we further approximate \eqref{r_k1} by using the following relationship \cite{6816003}
\begin{equation}\label{result}
	\mathbb{E}\left\{ {{{\log }_2}\left( {1 + \frac{X}{Y}} \right)} \right\} \approx {\log _2}\left( {1 + \frac{{\mathbb{E}\left\{ X \right\}}}{{\mathbb{E}\left\{ Y \right\}}}} \right),
\end{equation}
where $X$ and $Y$ are both non-negative random variables, but they are not required to be mutually independent. By applying \eqref{result} to \eqref{r_k1}, we obtain the following approximation
\begin{equation}\label{r_k2}
	{R_k} \approx {\log _2}\left(1 + \frac{{{\rho _{\rm d}}\mathbb{E}\left\{ {{{\left| {{{\hat{\bar{a}}}_{kk}}} \right|}^2}} \right\}}}{{{\rho _{\rm d}}\mathbb{E}\left\{ {{{\left| {{{\tilde{\bar{a}}}_{kk}}} \right|}^2}} \right\} + {\rho _{\rm d}}\sum\limits_{{k'} \ne k}^K {\mathbb{E}\left\{ {{{\left| {{a_{k{k'}}}} \right|}^2}} \right\} + 1} }}\right).
\end{equation}

After deriving the expectations in \eqref{r_k2}, the achievable downlink rate of the $k$-th user in the presence of downlink PSAs can be obtained, which is shown in \eqref{fr} on the top of this page, where
\begin{equation*}
	\begin{aligned}
		A &= {\varepsilon ^2}{\left({\tau _{\rm d}}{\rho _{\rm dp}}\xi_k  + 1\right)^2} + {\tau _{\rm d}^2}{\rho _{\rm dp}^2}{\xi_k ^3} + {\tau _{\rm d}}{\rho _{\rm dp}}{\xi_k ^2},\\
		B &= \xi_k  + {\tau _{\rm d}}{\rho _{\rm dp}}{\xi_k ^2} + \left(\sum\limits_{k \ne {k'}}^K {\mathbb{E}\{ {{\left| {{a_{k{k'}}}} \right|}^2}\}  + \frac{1}{{{\rho _{\rm d}}}}} \right){\left({\tau _{\rm d}}{\rho _{\rm dp}}\xi_k  + 1\right)^2},\\
		C &= 2{\tau _{\rm d}} \varepsilon_k \xi_k\sqrt {{\rho _{\rm dp}}{\mu _{\rm dp}}} \left({\tau _{\rm d}}{\rho _{\rm dp}}\xi_k  + 1\right),\\
		D &= {\tau _{\rm d}^2}{\rho _{\rm dp}}{\mu _{\rm dp}}{\xi_k ^2},\quad \mathbb{E}\{ {\left| {{a_{k{k'}}}} \right|^2}\}  = \sum\limits_{m = 1}^M {{\eta _{m{k'}}}} {\beta _{mk}}{\gamma _{m{k'}}},\\
		\varepsilon_k & = \sum\limits_{m = 1}^M {\sqrt {{\eta _{mk}}} } {\gamma _{mk}},\quad \xi_k  = \sum\limits_{m = 1}^M {{\eta _{mk}}} {\beta _{mk}}{\gamma _{mk}},
	\end{aligned}
\end{equation*}
and $\mathrm{SINR}''_{k}$ indicates the SINR of user $k$. The derivation of \eqref{fr} is detailed in Appendix \ref{app1}.

{\it Remark 3:} For the achievable downlink rate, it is observed that the transmit SNR for payload symbols $\rho_{\rm d}$ only exists in $B$. As $\rho_{\rm d}$ increases, the term $1/\rho_{\rm d}$ quickly becomes negligible, proving that $R_k$ is independent of $\rho_{\rm d}$ in this situation. Increasing the downlink transmit power, therefore,  does not help mitigate the effect of the downlink PSA. The observation that users use $\hat{{\bar{a}}}_{kk}$, which has already been tainted by the downlink PSAs, to decode the payload symbols can be used to explain this result.

{\it Remark 4:} $\rho_{\rm dp}$ and $\mu_{\rm dp}$ are two key parameters relating to legitimate APs and adversarial APs, respectively. The intuition behind is that the larger $\rho_{\rm dp}$ is, the greater $R_k$ will be. This can be confirmed by noting that the numerator of $\mathrm{SINR}''_{k}$ in \eqref{fr} is proportional to $\rho_{\rm dp}^2$, while its denominator is proportional to $\rho_{\rm dp}$. On the other hand, \eqref{fr} can be transformed into
\begin{equation}\label{sr}
\setcounter{equation}{27}	
	R_k \approx {\log _2}\left(2 + \frac{{C\sum\limits_{n = 1}^N {\sqrt {{\zeta _{nk}}} } {\kappa _{nk}} + A - B}}{{\mathcal{D}} + B}\right).
\end{equation}
Since $A$ and $B$ are independent of $\mu_{\rm dp}$, $\mathcal{D}$ is proportional to $\mu_{\rm dp}$ and $C$ is proportional to $\sqrt{\mu_{\rm dp}}$, thus (27) shows that increasing $\mu_{\rm dp}$ can reduce the achievable downlink rate. Also can be observed from (27), when $N$ is sufficiently large, it holds that $\sum\nolimits_{n = 1}^N {\sqrt {{\zeta _{nk}}} }{\kappa _{nk}} \approx N\mathbb{E}\left\{{\sqrt {{\zeta _{nk}}} }{\kappa _{nk}}\right\}$, where the expectation is taken with respect to $\theta_{nk}$. Since $\mathcal{D}$ is proportional to $N^2$, then the achievable downlink rate is a decreasing function of $N$. The rationale is straightforward, the greater the number of adversarial APs is, the lower the achievable downlink rate will be.

\subsection{Power Allocation from the Perspective of Adversarial APs}
The downlink PSA's ability to dramatically lower the achievable downlink rate has been proven. Additionally, adversarial APs can collude to lower the system's maximum achievable downlink rate by optimizing the power allocation parameters during the downlink training phase.

Since minimizing the maximum of $R_k$ is equivalent to minimizing the maximum of $\mathrm{SINR}''_k$, we utilize \eqref{fr} to construct the min-max optimization problem, i.e.,
\begin{equation}
	\begin{aligned}		
		{\bf OP1}:\quad &\mathop {\min }\limits_{\left\{ {{\zeta _{nk}}} \right\}} \mathop {\max }\limits_k ~\mathrm{SINR}''_{k}\\
		\mathrm{s.t.:}\quad&\sum\limits_{k = 1}^K {{\zeta _{nk}}} {\kappa _{nk}} \le 1,\quad\forall n\\
		&{\zeta _{nk}} \ge 0,\quad \forall n,\forall k
	\end{aligned}
\end{equation}
Let's define $\nu_{nk} = \sqrt{\zeta_{nk}}$, then {\bf OP1} can be transformed into
\begin{equation}\label{problemv2}
		\begin{aligned}		
		{\bf OP1.1}:\quad&\mathop {\min }\limits_{\left\{ {{\zeta _{nk}}} \right\}} \mathop {\max }\limits_k~ \frac{{C\sum\limits_{n = 1}^N {{\nu _{nk}}} {\kappa _{nk}} + \mathcal{D} + A}}{{\mathcal{D} + B}}\\
		\mathrm{s.t.:}\quad&\sum\limits_{k = 1}^K {{\nu^2 _{nk}}} {\kappa _{nk}} \le 1,\quad\forall n\\
		&{\nu _{nk}} \ge 0,\quad \forall n,\forall k
	\end{aligned}
\end{equation}

As {\bf OP1.1} is quasiconcave, the bisection method can be used to resolve this problem. Towards this end, we first formulate the following equivalent problem by introducing an auxiliary variable $t$, i.e.,
\begin{equation}\label{prblemv3}
	\begin{aligned}
		{\bf OP1.2}:\quad &\mathop {\min }\limits_{\left\{ {{\nu _{nk}}},t \right\}} ~t\\
		\mathrm{s.t.:}\quad &{\left\| {{{\bm{\nu }}_k^T}{{\bm{\kappa }}_k} + \frac{C}{{2D}}\sqrt {\frac{A}{D}} } \right\|^2} \le \left(t - 1\right){\left\| {{{\bm{\nu }}_k}\circ{{{\bar{\bm\kappa }}}_k}\circ{{\bar {\bm{\theta}} }_k}} \right\|^2} + \\
		\qquad &\qquad \qquad \qquad \qquad \quad \quad t{\left({{\bm{\nu }}_k^T}{{\bm{\kappa }}_k}\right)^2} + \frac{{Bt}}{D} + \frac{{{C^2}}}{{4{D^2}}}\\
	&	\sum\limits_{k = 1}^K {{\nu _{nk}^2}} {\kappa _{nk}} \le 1, \quad\forall n\\
		&{\nu _{nk}} \ge 0,\quad \forall n, \forall k
	\end{aligned}
\end{equation}
where
\begin{equation*}
	\begin{aligned}
		{\bm\kappa }_k &= {\left[ {{\kappa _{1k}},{\kappa _{2k}},\dots,{\kappa _{Nk}}} \right]^T},\\
		\bar{\bm\kappa }_k &= {\left[ {\sqrt {{\kappa _{1k}}} ,\sqrt {{\kappa _{2k}}},\dots,\sqrt {{\kappa _{Nk}}} } \right]^T},\\
		{\bm\theta }_k &= {\left[ {{\theta _{1k}},{\theta _{2k}},\dots,{\theta _{Nk}}} \right]^T},\\
		\bar{\bm\theta }_k &= {\left[ {\sqrt {{\theta _{1{\rm{ }}k}}},\sqrt {{\theta _{2k}}},\dots,\sqrt {{\theta _{Nk}}} } \right]^T},\\
		{\bm\nu }_k &= {\left[ {{\nu _{1k}},{\nu _{2k}},\dots,{\nu _{Nk}}} \right]^T},
	\end{aligned}
\end{equation*}
and ``$\circ$'' is the dot-product operator.

Since the first constraint in \eqref{prblemv3} is neither convex or concave with respect to $\nu_{nk}$, we resort to the sequential convex approximation (SCA) method to overcome such non-convexity. For ease of exposition, we define the left-hand side of the first constraint in \eqref{prblemv3} as $f\left(\bm{\nu}_k\right)$. Then according to the SCA concept, $f\left(\bm{\nu}_k\right)$ can be expanded by using the first-order Taylor expansion, i.e.,
\begin{equation}
	\begin{aligned}
		\hat f\left({{\bm{\nu }}_{{k}}};{\bm{\nu }}_{{k}}^{{n}}\right) =~& f\left({\bm{\nu }}_{{k}}^{{n}}\right) + \left({{\bm{\nu }}_{{k}}} - {\bm{\nu }}_{{k}}^{{n}}\right)^T\nabla f\left({\bm{\nu }}_{{k}}^{{n}}\right)\\
		=~&\left(t - 1\right){\left\| {{{\bm{\nu }}_{{k}}}\circ{\bar{\bm\kappa }_k}\circ\bar{\bm\theta }_k} \right\|^2} + 2t{\bm{\nu }}_{{k}}^T{{\bm{\kappa }}_{{k}}}{{\bm\kappa }}_{{k}}^T{\bm{\nu }}_{{k}}^{{n}}\\
		& + 2\left(t - 1\right){\left({{\bm{\nu }}_{{k}}} - {\bm{\nu }}_{{k}}^{{n}}\right)^T}({\bm{\nu }}_{{k}}^{{n}}\circ{{\bm{\kappa }}_{{k}}}\circ{\bm\theta _{{k}}})\\
		& - t{\left\| {{\bm{\kappa }}_{{k}}^T{\bm{\nu }}_{{k}}^{{n}}} \right\|^2} + \frac{{tB}}{D} + \frac{{{C^2}}}{{4{D^2}}},
	\end{aligned}
\end{equation}
where ${\bm{\nu}_k^n} = \left[\nu_{1k}^n,\nu_{2k}^n,\dots,\nu_{Nk}^n\right]^T$ denotes the value of $\bm{\nu}_k$ for the $n$-th iteration. Therefore, the first constraint in \eqref{prblemv3} can be modified into
\begin{equation}
	\begin{aligned}
			\left\| {{{\bm{\nu }}_{{k}}^T}{{\bm{\kappa }}_{{k}}} + \frac{C}{{2D}}\sqrt {\frac{A}{D}}} \right\|^2 \le \left(t - 1\right){\left\| {{{\bm{\nu }}_{{k}}}\circ{{{\bar{\bm \kappa }}}_{{k}}}\circ{{\bar {\bm \theta} }_{{k}}}} \right\|^2}\\
			+ 2\left(t - 1\right){\left({{\bm{\nu }}_{{k}}} - {\bm{\nu }}_{{k}}^{{n}}\right)^T}\left({\bm{\nu }}_{{k}}^{{n}}\circ{{\bm{\kappa }}_{{k}}}\circ{\bm \theta _{{k}}}\right)\\
			+ 2t{\bm{\nu }}_{{k}}^T{{\bm{\kappa }}_{{k}}}{\bm{\kappa }}_{{k}}^T{\bm{\nu }}_{{k}}^{{n}}- t{\left\| {{\bm{\kappa }}_{{k}}^T{\bm{\nu }}_{{k}}^{{n}}} \right\|^2} + \frac{{tB}}{D} + \frac{{{C^2}}}{{4{D^2}}}.
	\end{aligned}
\end{equation}
The above inequality can be further simplified as
\begin{equation}\label{constraint1}
	\begin{aligned}
		{\left\| {{{\bf{V}}_k}} \right\|^2}&\le t{\left\| {{{\bm{\nu }}_{{k}}}\circ{{{\bar{\bm \kappa }}}_{{k}}}\circ{{\bar {\bm \theta} }_{{k}}}} \right\|^2}+2t{\bm{\nu }}_{{k}}^T{{\bm{\kappa }}_{{k}}}{\bm{\kappa }}_{{k}}^T{\bm{\nu }}_{{k}}^{{n}}\\
		&+ 2\left(t - 1\right){\left({{\bm{\nu }}_{{k}}} - {\bm{\nu }}_{{k}}^{{n}}\right)^T}\left({\bm{\nu }}_{{k}}^{{n}}\circ{{\bm{\kappa }}_{{k}}}\circ{\bm \theta _{{k}}}\right)+\frac{{tB}}{D}
	\end{aligned}
\end{equation}
where
\begin{equation*}
	{\bf{V}}_k = \left[{{\bm{\kappa }}_{{k}}^T{\bm{\nu }}_{{k}}} + {\frac{C}{{2D}}\sqrt {\frac{A}{D} - \frac{C}{{4{D^2}}}} } \quad \sqrt{t}\bm{\kappa}_k^T\bm{\nu}_k^n\quad {{{\bm{\nu }}_{{k}}}\circ{{{\bar{\bm \kappa }}}_{{k}}}\circ{{\bar {\bm \theta} }_{{k}}}}\right].
\end{equation*}

%\begin{equation}
%	\begin{aligned}
%		{\left\| {{\bf{Ax + b}}} \right\|^2} \le v({{\bf{c}}^{{T}}}{\bf{x}} + d)\\
%		{{\bf{c}}^{{T}}}{\bf{x}} + d \ge 0\\
%		v \ge 0
%	\end{aligned}
%\end{equation}
%\begin{equation}
%	\left\| {\left[ \begin{aligned}
%			2({\bf{Ax + b}})\\
%			{{\bf{c}}^{{T}}}{\bf{x}} + d - v
%		\end{aligned} \right]} \right\| \le {{\bf{c}}^{{T}}}{\bf{x}} + d + v
%\end{equation}

The inequality in \eqref{constraint1} describes a hyperbolic constraint and represents a class of convex problems that can be casted as second-order cone programs (SOCPs). By virtual of properties of hyperbolic constraints \cite{LOBO1998193}, i.e.,
\begin{equation}
	w^2 \le xy,~x \ge 0,~y \ge 0 \Leftrightarrow \left\| {\left[ {\begin{array}{*{20}{c}}
				x\\
				y
		\end{array}} \right]} \right\| \le x + y,
\end{equation}
where $x$, $y$, and $w$ are the optimization
variables or parameters, {\bf OP1.2} is then reconstructed as
\begin{equation}
	\begin{aligned}		
		{\bf OP1.3}:\quad &\mathop {\min }\limits_{\left\{ {{\zeta _{nk}}},t \right\}}  ~t\\
		\mathrm{s.t.:}\quad &\left\|\left[\bm{\mathcal{S}}_k ~ S_k-t\right]\right\| \le S_k + t \\
		& S_k \ge 0, \quad \forall k\\
		&\sum\limits_{k = 1}^K {{\nu^2 _{nk}}} {\kappa _{nk}} \le 1,\quad\forall n\\
		&{\nu _{nk}} \ge 0,\quad \forall n,\forall k
	\end{aligned}
\end{equation}
where
\begin{equation*}
	\small
	\bm{\mathcal{S}}_k = \left[2{{\bm{\kappa }}_{{k}}^T{\bm{\nu }}_{{k}}}+ {\frac{2C}{{D}}\sqrt {\frac{A}{D} - \frac{C}{{4{D^2}}}} } \quad 2{{{\bm{\nu }}_{{k}}}\circ{{{\bar{\bm \kappa }}}_{{k}}}\circ{{\bar {\bm \theta} }_{{k}}}}\quad 2\sqrt{t}\bm{\nu}_k^{nT}\bm{\kappa}_k^T\right].
\end{equation*}
For a fixed $t$, the constraints in {\bf OP1.3} are convex and it is possible to determine whether a given $t$ is feasible or not. Hence we can apply the bisection method as follows. In particular, we first choose an interval $\left(t_{\min},t_{\max}\right)$ that contains the optimal value $t^*$. After that, we check the feasibility of the midpoint $t = \left(t_{\min} + t_{\max}\right)/2$. If $t$ is feasible, the search interval is updated to $\left(t,t_{\max}\right)$, or otherwise  $\left(t_{\min},t\right)$. This process continues until the maximum number of iterations is reached or the search interval becomes small enough. The optimal power allocation coefficients are then inserted into \eqref{fr} to evaluate the actual system performance.

From the above discussions, uplink channel estimation is key to successfully launching downlink PSAs. Therefore, the impact of downlink PSAs can be significantly reduced if one can stop adversarial APs from estimating the uplink channel. This is a difficult task in general, since by definition adversarial APs are passive and never transmit during uplink training phase. 

\section{Attacking the Downlink Data Transmission}\label{IIIII}
Aside from the PSAs in the downlink training phase, adversarial APs have another means of interfering with legitimate communications. For example, they precode random noise and transmit it to users during the downlink data transmission phase. Since there are many studies focusing on the detection of and countermeasure to the PSA, targeting the downlink data transmission may provide an alternative to the PSA from the standpoint of adversarial APs. 
\subsection{Achievable Rate Analysis}
We assume that adversarial APs obtain their channels to users in the uplink training phase and remain silent during the downlink training phase, so the estimation of the effective channel $a_{kk}$ is free of malicious attacks. In particular, the precoded signal by the $n$-th adversarial AP is expressed as
\begin{equation}\label{scenario2}
	{x_{{\rm da},n}} = \sqrt {\mu_{\rm da}} \sum\limits_{k = 1}^K {\sqrt {{\zeta _{nk}}} {{\hat f}^*}_{nk}{j_k}},
\end{equation}
where the subscript ``da'' denotes downlink signals sent from adversarial APs, $\mu_{\rm da}$ is the normalized transmit SNR during the downlink data phase, and $j_k$ denotes the randomly generated noise intended for the $k$-th user with $\mathbb{E}\small\{ {\left| j_k \right|^2} \small\} = 1$. As can be seen from  \eqref{scenario2}, the power constraint of the adversarial APs is given by
\begin{equation}
	\mathbb{E}\left\{ {{{\left| {x_{{\rm da},n}} \right|}^2}} \right\} \le  \rho_{\rm da} \quad {\rm or} \quad \sum\limits_{k = 1}^K {{\zeta _{nk}}{\kappa _{nk}}}  \le 1.
\end{equation}
Then in the downlink transmission, the received signal by the $k$-th user is expressed as
\begin{equation}\label{dl_psa}
	\begin{aligned}
		&{r_{{\rm da},k}} = \sum\limits_{m = 1}^M {{g_{mk}}{x_{{\rm d},m}}}  + \sum\limits_{n = 1}^N {{f_{nk}}{x_{{\rm da},n}}}  + {w_{{\rm da},k}}\\
		&=\sqrt {\rho_{\rm d}} {a_{kk}}{q_k} + \sqrt {\rho_{\rm d}} \sum\limits_{{k'} \ne k}^K {{a_{k{k'}}}{q_{k'}}} + \sqrt {\rho _{\rm da}} \sum\limits_{k'=0}^K b_{kk'}j_{k'} + {w_{{\rm da},k}},
	\end{aligned}
\end{equation}
where $x_{{\rm d},m}$ is shown in \eqref{pcoding-data}. We use symbol ${r_{{\rm da},k}}$ to represent the received signal in the presence of an attack during the downlink data transmission.

{\it Remark 5:} It is shown that the first, second, and third components of \eqref{dl_psa} represent the desired signal, inter-user interference, and random noise transmitted by adversarial APs, respectively. Obviously, the jamming signal $j_k$ will reduce the effective SINR of ${r_{{\rm da},k}}$, and finally degrade the achievable downlink rate. An ensuing question is which strategy, namely the downlink PSA in \eqref{eve_precoding} or downlink random noise in \eqref{scenario2}, is more detrimental to the achievable rate of legitimate communications given the same transmit power budget.

As before, the received signal is rewritten as 
\begin{equation}
	{r_{{\rm da},k}} = \sqrt {{\rho _{\rm d}}} {a_{kk}}{s_k} + {\tilde w_{{\rm da},k}},
\end{equation}
where ${\tilde w_{{\rm da},k}} = \sqrt {\rho_{\rm d}} \sum\nolimits_{{k'} \ne k}^K {{a_{k{k'}}}{q_{k'}}} + \sqrt {\rho _{\rm da}} \sum\nolimits_{k'=0}^K b_{kk'}j_{k'} + {w_{{\rm da},k}}$ denotes the effective noise in this case. We assume $q_k$ and $j_k$ are mutually uncorrelated without loss of generality. Then following the same approach in \eqref{zero_corre}, a lower bound of the per-user achievable rate is computed by \eqref{fr1}, shown on the top of next page. Note that we use $R'_k$ to symbolize the per-user achievable rate in the scenario of attacking downlink data transmission. By applying the approximation in \eqref{gs-appr}, \eqref{fr1} can be transformed into
\begin{equation}
	\setcounter{equation}{41}
	R'_k = \mathbb{E}\left\{\log_2\left(1+ \lambda_{k} \right)\right\}
\end{equation}
where 
\begin{equation*}
	\lambda_{k} = \frac{\rho_{\rm d}\left|\hat{{a}}_{kk}\right|^2}{\rho_{\rm d}\mathbb{E}\small\{\left|\tilde{a}_{kk}\right|^2\small\} +  \rho_{\rm d}\sum\limits_{k'\ne k}^K\mathbb{E}\small\{\left|a_{kk'}\right|^2|\hat{a}_{kk}\small\} + \rho_{\rm da}\left|{b}_{kk}\right|^2 + 1}.
\end{equation*}
Since the uplink training is free of PSAs, the estimated channel by using \eqref{dl-channel-estimate} is employed here. 

Due to the presence of the conditional probability, it is challenging to derive a closed-form expression for (40). We resort to \eqref{result} to simplify $R'_k$, i.e.,
\begin{equation}\label{rate_data_1}
	\begin{aligned}
		R'_k \approx \log_2\left(1 + \lambda'_k\right),
	\end{aligned}
\end{equation}
where
\begin{equation*}
	\lambda'_k = \frac{\rho_{\rm d}\mathbb{E}\small\{\left|\hat{a}_{kk}\right|^2\small\}}{\rho_{\rm d}\mathbb{E}\small\{\left|\tilde{a}_{kk}\right|^2\small\} + \rho_{\rm d}\sum\limits_{k'\ne k}^{K}\mathbb{E}\small\{\left|{a}_{kk'}\right|^2\small\} +\rho_{\rm da}\mathbb{E}\small\{\left|{b}_{kk}\right|^2\small\} + 1}.
\end{equation*}
The numerator in $\lambda'_{k}$ denotes the beamforming gain, while the first to fourth components of the denominator indicate the variance of the channel estimation error, the multi-user interference, the interference caused by adversarial APs, and the additive noise, respectively. The per-user achievable rate in this case is found by calculating the expectations in \eqref{rate_data_1}, of which the detailed derivation is described in Appendix \ref{app2}. The final result is shown in \eqref{fr2} at the top of this page, where $\varrho_{k} = \frac{\tau_{\mathrm{dp}}\rho_{\rm dp}\varsigma_{kk}^2}{1 + \tau_{\mathrm{dp}}\rho_{\mathrm{dp}}\varsigma_{kk}}$, and $\lambda''_{k}$ denotes the SINR.

{\it Remark 6:} Adversarial APs can change the transmit power to affect the achievable downlink rate since $\rho_{\rm da}$ is included in the denominator of $\lambda''_{k}$. In addition, $\rho_{\rm d}$ and $\rho_{\rm da}$ interact together to determine the ultimate achievable rate. To interfere with legitimate communications, adversarial APs can choose to boost the transmit power or optimize the power allocation given a fixed power budget. Although the first method is straightforward, it may make it easier for legitimate nodes to identify an attack. In terms of improving the hiding capability of adversarial APs, the second method is preferable.

\begin{figure*}[!t]	
	\setcounter{mytempeqncnt}{\value{equation}}	
	\setcounter{equation}{39}
	\begin{equation}\label{fr1}
		R'_k \ge \mathbb{E}\left\{\log_2\left(1 + \frac{\rho_{\rm d}\left|\mathbb{E}\left\{a_{kk}|\hat{a}_{kk}\right\}\right|^2}{\rho_{\rm d}\sum\limits_{{k'} =1}^K\mathbb{E}\left\{\left|a_{kk'}\right|^2|\hat{{a}}_{kk}\right\} + \rho_{\rm d}\sum\limits_{{k'} =1}^K\mathbb{E}\left\{\left|b_{kk'}\right|^2|\hat{{a}}_{kk}\right\} - \rho_{\rm da}\left|\mathbb{E}\left\{a_{kk}|\hat{a}_{kk}\right\}\right|^2 + 1}\right)\right\}
	\end{equation}
	\setcounter{equation}{\value{mytempeqncnt}}	
	\hrulefill	
	\vspace*{4pt}	
\end{figure*}
\begin{figure*}[!t]	
	\setcounter{mytempeqncnt}{\value{equation}}	
	\setcounter{equation}{41}
	\begin{equation}\label{fr2}
		R_k = \log_2\left(1+\lambda''_{k}\right) = \log_2\left(1 + \frac{\rho_{\rm d}\left(\sum\limits_{m=1}^{M}\sqrt{\eta_{mk}}\gamma_{mk}\right)^2 + \rho_{\rm d}\varrho_{{k}}} {\rho_{\rm d}\sum\limits_{k'=1}^{K}\varsigma_{kk'} - \rho_{\rm d}\varrho_k + \underbrace{\rho_\mathrm{da}\left(\sum\limits_{n=1}^{N}\sqrt{\zeta_{nk}}\kappa_{nk}\right)^2 + \rho_{\rm da} \sum\limits_{k'=1}^K\sum\limits_{n=1}^{N}\zeta_{nk'}\theta_{nk}\kappa_{nk'}}_{\mathcal{F}} + 1} \right)
	\end{equation}
	\setcounter{equation}{\value{mytempeqncnt}}	
	\hrulefill	
	\vspace*{4pt}	
\end{figure*}
\subsection{Power Allocation}
As in the case of downlink PSAs, we investigate the optimal power allocation towards $\zeta_{nk}$ by using the min-max criterion. Specifically, because of the relationship between the rate and SINR, the expression of $\lambda''_{k}$ in \eqref{fr2} is considered. As a result, the optimization problem can be formulated as
\begin{equation}
	\setcounter{equation}{43}
	\begin{aligned}		
		{\bf OP2}:\quad &\mathop {\min }\limits_{\left\{ {{\zeta _{nk}}} \right\}} \mathop {\max }\limits_k ~ \lambda''_{k}\\
		\mathrm{s.t.:}\quad&\sum\limits_{k = 1}^K {{\zeta _{nk}}} {\kappa _{nk}} \le 1,\quad\forall n\\
		&{\zeta _{nk}} \ge 0,\quad \forall n,\forall k
	\end{aligned}
\end{equation}
Unfortunately, {\bf OP2} is neither convex or concave with respect to $\zeta_{nk}$, thus it cannot be solved directly. Given $\tau_{\mathrm{dp}} \ge K$, the following approximation can be assumed, i.e.,
\begin{equation}
	\varrho_{k} = \frac{\tau_{\mathrm{d}}\rho_{\rm dp}\varsigma_{kk}^2}{1 + \tau_{\mathrm{d}}\rho_{\mathrm{dp}}\varsigma_{kk}} \approx \varsigma_{kk},\quad \text{if}\quad \tau_{\mathrm{d}}\rho_{\mathrm{dp}}\varsigma_{kk} \gg 1.
\end{equation}
The condition $\tau_{\mathrm{d}}\rho_{\mathrm{dp}}\varsigma_{kk} \gg 1$ is easily satisfied because $\rho_{\mathrm{dp}} \ge 1$. Using this approximation, {\bf OP2} can be converted into
\begin{equation}\label{problemv2.1}
	\begin{aligned}		
		{\bf OP2.1}:\quad &\mathop {\min }\limits_{\left\{ {{\zeta _{nk}}} \right\}} \mathop {\max }\limits_k~ \frac{\rho_{\rm d}\left(\sum\limits_{m=1}^{M}\sqrt{\eta_{mk}}\gamma_{mk}\right)^2 + \rho_{\rm d}\varsigma_{kk}}{\rho_{\rm d}\sum\limits_{k'\ne k}^K \varsigma_{kk'}  + \mathcal{F} + 1}\\
		\mathrm{s.t.:}\quad&\sum\limits_{k = 1}^K {{\zeta _{nk}}} {\kappa _{nk}} \le 1,\quad\forall n\\
		&{\zeta _{nk}} \ge 0,\quad \forall n,\forall k
	\end{aligned}
\end{equation}
where the expression  of $\mathcal{F}$ is shown in \eqref{fr2}. Defining $\nu_{nk} = \sqrt{\zeta_{nk}}$, then after some mathematical manipulations, the optimization problem {\bf OP2.1} can be further transformed into
\begin{equation}\label{prblemv3.2}
	\begin{aligned}
		{\bf OP2.2}:\quad &\mathop {\min }\limits_{\left\{ {{\nu _{nk}}} \right\},t} ~t\\
		\mathrm{s.t.:}\quad &\frac{\rho_{\rm d}}{t}\left\|\bm{v}_k\right\|^2 \le \mathcal{F} + \rho_{\rm da}\sum_{n=1}^{N} \theta_{nk}\kappa_{nk}\nu_{nk}^2 + \\
		& \qquad \rho_{\rm d}\sum_{k' \ne k}^{K}\sum_{m=1}^{M} \eta_{mk'}\beta_{mk}\gamma_{mk'} +1\\
		&	\sum\limits_{k = 1}^K \kappa_{nk}\nu_{nk}^2 \le 1, \quad \forall n,\\
		&{\nu _{nk}} \ge 0,\quad \forall n, \forall k
	\end{aligned}
\end{equation}
where
\begin{equation*}
	\begin{aligned}
		\bm{v}_k &= \left[\bm{v}_{k1}^T~\bm{v}_{k2}^T\right]^T,\\
		\bm{v}_{k1} &= \left[\sqrt{\zeta_{1k}}\kappa_{1k},\dots,\sqrt{\zeta_{\textsc{N}k}}\kappa_{Nk}\right]^T,\\
		\bm{v}_{k2} &= \left[\sqrt{\zeta_{1k}\theta_{1k}\kappa_{1k}},\dots,\sqrt{\zeta_{Nk}\theta_{Nk}\kappa_{Nk}}\right]^T.
	\end{aligned}
\end{equation*}

%\begin{equation}
%	\begin{aligned}
%		\bm{\kappa}_k & = \left[\kappa_{1k},\dots,\kappa_{Nk}\right]^T,\\
%		{\bar{\bm\kappa}_k} & = \left[\sqrt{\kappa_{1k}},\dots,\sqrt{\kappa_{Nk}}\right]^T,\\
%		\bm{\theta}_k & = \left[\theta_{1k},\dots,\theta_{Nk}\right]^T,\\
%		{\bar{\bm\theta}_k} & = \left[\sqrt{\theta_{1k}},\dots,\sqrt{\theta_{Nk}}\right]^T.
%	\end{aligned}
%\end{equation}
As the first constraint in {\bf OP2.2} is neither convex or concave, we turn to the SCA to deal with such non-convexity. We use the first-order Taylor expansion as the first step to obtain a convex approximation of function $f\left(\bm{\nu}_k\right) = \left(\bm{\kappa}_k^T\bm{\nu}_k\right)^2 + \left\|{\bar{\bm\kappa}_k}\circ {\bar{\bm\theta}_k} \circ \bm{\nu}_k\right\|^2$, where $\bm{\kappa}_k$, ${\bar{\bm\kappa}_k}$, $\bm{\nu}_k$, and $\bar{\bm\theta}_k$ are defined in \eqref{prblemv3}. Denoted by $\bm{\nu}_{k}^n$ the value of $\bm{\nu}_{k}$ for the $n$-th iteration of SCA, where $\bm{\nu}_k^n = \left[\nu^n_{1k},\dots,\nu_{Nk}^n\right]^T$. As a result, $f\left(\bm{\nu}_k\right)$ can be approximated as
\begin{equation}
	\begin{aligned}
		\hat{f}\left(\bm{\nu}_k;\bm{\nu}_k^n\right) & = f\left(\bm{\nu}_{k}^n\right) + \left(\bm{\nu}_{k}-\bm{\nu}_{k}^n\right)^T \nabla f\left(\bm{\nu}_{k}^n\right)\\
		& = \left\|{\bar{\bm\kappa}_k}\circ {\bar{\bm\theta}_k} \circ \bm{\nu}_k^n\right\|^2 + 2\bm{\nu}_k^T\bm{\kappa}_k\bm{\kappa}_k^T\bm{\nu}_k^n - \left\|\bm{\kappa}_k^T\bm{\nu}_k^n\right\|^2\\
		&+2\left(\bm{\nu}_{k}-\bm{\nu}_{k}^n\right)^T\left({\bm{\kappa}_k}\circ {\bm{\theta}_k} \circ \bm{\nu}_k^n\right),
	\end{aligned}
\end{equation}
where $\hat{f}\left(\bm{\nu}_k;\bm{\nu}_k^n\right)$ denotes the approximation of $f\left(\bm{\nu}_k\right)$ after $n$-th iteration. Therefore, the first constraint in {\bf OP2.2} can be rewritten as
\begin{equation}\label{48}
	\begin{aligned}
		&\frac{\rho_{\rm d}}{\rho_{\rm da}t}\left\|\bm{v}_k\right\|^2 \le \left\|{\bar{\bm\kappa}_k}\circ {\bar{\bm\theta}_k} \circ \bm{\nu}_k^n\right\|^2 + 2\bm{\nu}_k^T\bm{\kappa}_k\bm{\kappa}_k^T\bm{\nu}_k^n - \left\|\bm{\kappa}_k^T\bm{\nu}_k^n\right\|^2\\
		&+ 2\left(\bm{\nu}_{k}-\bm{\nu}_{k}^n\right)^T\left({\bm{\kappa}_k}\circ {\bm{\theta}_k} \circ \bm{\nu}_k^n\right)  + \frac{1}{\rho_{\rm da}}\\
		&+ \sum_{k'\ne k}^{K}\sum_{n=1}^{N} \zeta_{nk'}\theta_{nk}\kappa_{nk'} + \frac{\rho_{\rm d}}{\rho_{\rm da}}\sum_{k' \ne k}^{K}\sum_{m=1}^{M} \eta_{mk'}\beta_{mk}\gamma_{mk'}.
	\end{aligned}
\end{equation}
We observe that \eqref{48} equals
\begin{equation}\label{constr.1}
	\begin{aligned}
		&{\left\|\bar{\bm{v}}_k\right\|^2} \le t\left(\rho_{\rm da}\left\|{\bar{\bm\kappa}_k}\circ {\bar{\bm\beta}_k} \circ \bm{\nu}_k^n\right\|^2 + 2\rho_{\rm da}\bm{\nu}_k^T\bm{\kappa}_k\bm{\kappa}_k^T\bm{\zeta}_k^n\right.\\
		&+ \rho_{\rm da}\sum_{k'\ne k}^{K}\sum_{n=1}^{N} \zeta_{nk'}\theta_{nk}\kappa_{nk'} + \rho_{\rm d}\sum_{k' \ne k}^{K}\sum_{m=1}^{M} \eta_{mk'}\beta_{mk}\gamma_{mk'}\\
		&\left.+ 2\rho_{\rm da}\left(\bm{\nu}_{k}-\bm{\nu}_{k}^n\right)^T\left({\bm{\kappa}_k}\circ {\bm{\theta}_k} \circ \bm{\nu}_k^n\right)  - \left\|\bm{\kappa}_k^T\bm{\nu}_k^n\right\|^2 + 1\right)
	\end{aligned}
\end{equation}
where $\bar{\bm{v}}_k = \left[\sqrt{\rho_{\rm d}}\bm{v}_{k1}^T~\sqrt{\rho_{\rm d}}\bm{v}_{k2}^T~\sqrt{t\rho_{\rm da}}\bm{\kappa}_k^T\bm{\nu}_{k}^n\right]$. We highlight that $\bar{\bm{v}}_k$ describes a linear constraint. Finally, one can arrive at the following optimization problem
\begin{equation}\label{prblemv3.1}
	\begin{aligned}
		{\bf OP2.3}\quad &\mathop {\min }\limits_{\left\{ {{\nu _{nk}}} \right\},t} ~t\\
		\mathrm{s.t.:}\quad & \mathrm{Eq}.~\eqref{constr.1}\\
		&	\sum\limits_{k = 1}^K \kappa_{nk}\nu_{nk}^2 \le 1, \quad \forall n,\\
		&{\nu _{nk}} \ge 0,\quad \forall n, \forall k
	\end{aligned}
\end{equation}
which is a convex optimization problem and can be solved efficiently by numerical methods, e.g., the bisection method.

So far, we have investigated two scenarios of attacks carried out by adversarial APs. The first scenario involves the downlink PSA, which contaminates the estimation results of the effective channels. The second one, inspired by the downlink PSA, attacks the downlink data transmission phase directly. While it is difficult to quantify the rate loss by comparing \eqref{fr1} to \eqref{fr2}, numerical simulation can effectively demonstrate which type of attacks can reduce the legitimate rate more within a given transmit power.

\section{Simulation Results and Discussions}\label{IIIIII}
In this part, numerical simulation results are presented to showcase the impact of downlink PSAs on the achievable rate of the cell-free mMIMO system. We assume $M$ legitimate APs, $N$ adversarial APs, and $K$ users are uniformly distributed at random within a square area of $\text{1~km}^2$. As for the wireless channel, the large-scale fading factor $\beta_{mk}$ models the path loss and shadowing fading, according to
\begin{equation}\label{ls-factor}
	\beta_{mk} = \text{PL}_{mk} 10^{\frac{\sigma_{\mathrm{sh}}z_{{mk}}}{10}},
\end{equation}
where $\text{PL}_{mk}$ indicates the path loss, $10^{\frac{\sigma_{\mathrm{sh}}z_{{mk}}}{10}}$ describes the shadow fading with a standard derivation  of $\sigma_{\mathrm{sh}}$, and $z_{mk} \sim \mathcal{N}\left(0,1\right)$. The modeling of $\theta_{nk}$ related to adversarial APs is the same as that in \eqref{ls-factor}. The small-scale fading factors follow a distribution of $\mathcal{CN}\left(0,1\right)$. Unless stated otherwise, the transmit power of adversarial APs in the downlink training phase is assumed to be the same as that of legitimate ones. Other simulation parameters used in our study are listed in Table \ref{table1}.

\begin{table}
	\centering
	\caption{The Simulation Parameters}
	\label{table1}
	\begin{IEEEeqnarraybox}[\IEEEeqnarraystrutmode\IEEEeqnarraystrutsizeadd{1pt}{1pt}]{v/c/v/c/v}		
		\IEEEeqnarrayrulerow\\
		\IEEEeqnarrayseprow[1pt]\\
		&\text{Number of legitimate APs ($M$)}&&128\\
		\IEEEeqnarrayseprow[1pt]\\
		\IEEEeqnarrayrulerow\\
		\IEEEeqnarrayseprow[1pt]\\
		&\text{Number of adversarial APs ($N$)}&&32\\
		\IEEEeqnarrayseprow[1pt]\\
		\IEEEeqnarrayrulerow\\
		\IEEEeqnarrayseprow[1pt]\\
		&\text{Number of users ($K$)}&&4\\
		\IEEEeqnarrayseprow[1pt]\\
		\IEEEeqnarrayrulerow\\
		\IEEEeqnarrayseprow[1pt]\\
		&\text{Bandwidth}&& 20~\rm MHz\\
		\IEEEeqnarrayseprow[1pt]\\
		\IEEEeqnarrayrulerow\\
		\IEEEeqnarrayseprow[1pt]\\
		&\text{Carrier frequency}&& 1.9~\rm GHz\\
		\IEEEeqnarrayseprow[1pt]\\
		\IEEEeqnarrayrulerow\\
		\IEEEeqnarrayseprow[1pt]\\
		&\text{Length of pilot sequences}&& 32\\
		\IEEEeqnarrayseprow[1pt]\\
		\IEEEeqnarrayrulerow\\
		\IEEEeqnarrayseprow[1pt]\\
		&\text{Transmit power of downlink pilots}&&200 ~\rm mW\\
		\IEEEeqnarrayseprow[1pt]\\
		\IEEEeqnarrayrulerow
	\end{IEEEeqnarraybox}
\end{table}
\subsection{Downlink Achievable Rate Analysis}
\begin{figure}
	\centering
	\includegraphics[width=85mm]{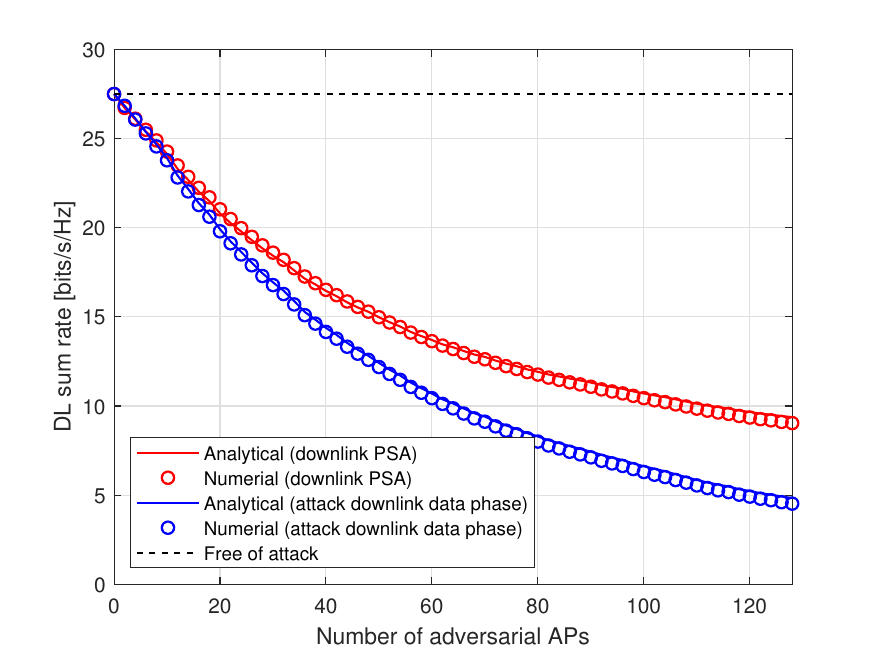}
	\caption{Downlink achievable sum rate versus the number of adversarial APs, where $M=128$ and $K=4$.}
	\label{fig2}
\end{figure}

Fig. \ref{fig2} shows how the achievable downlink sum rate varies with the number of adversarial APs $N$, where $M=128$ and $K=4$. The achievable downlink sum rate is defined as
\begin{equation}
	R_{\rm sum} = \alpha^{\rm{DL}}\left(1-\frac{\tau_{\rm u}+\tau_{\rm d}}{\tau_{\rm c}}\right)\sum\limits_{k=1}^K{R}_{k},
\end{equation}
where $\alpha^{\rm{DL}}$ is the fraction of data symbols that are used for downlink payload transmission, $\tau_{\rm d}$ is the length of downlink symbols, and $\tau_{\rm c}$ indicates the length of a coherence interval. We focus on three scenarios in particular, namely the attacks by adversarial APs on the downlink training and data transmission phases, as well as the absence of such attacks. First, the accuracy of our derivation is validated by the good agreement between our numerical and analytical results. It is interesting to note that, as shown by a rate difference between these two cases, targeting downlink data transmission is more successful than attacking the training phase. Additionally, a significant drop in the downlink attainable rate is observed as $N$ rises. Therefore, the need for efficient countermeasures is crucial. In particular, one of the solutions is to prevent adversarial APs from obtaining channel information during uplink training \cite{6288501}.

\begin{figure}
	\centering
	\includegraphics[width=85mm]{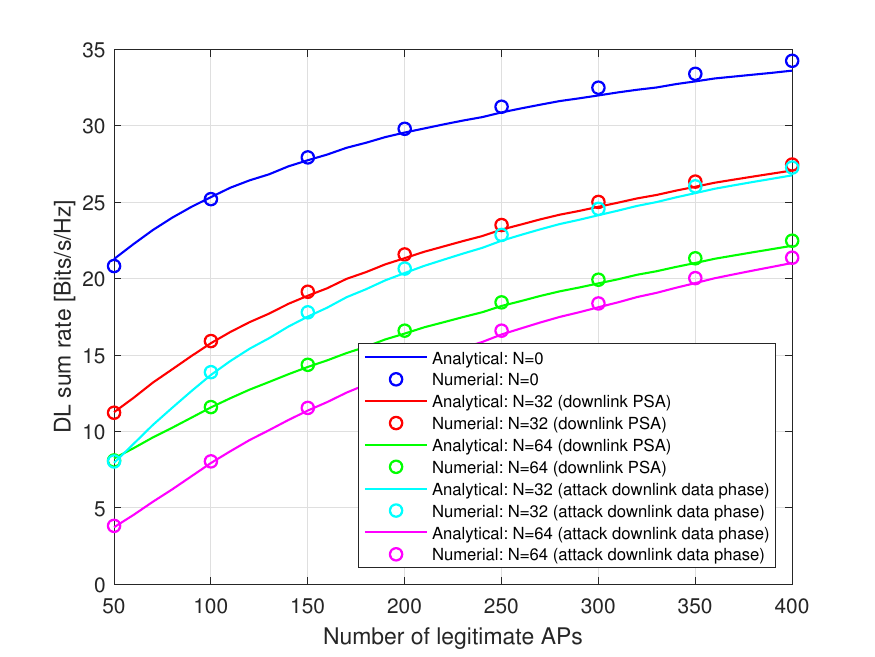}
	\caption{Downlink achievable sum rate versus the number of legitimate APs, where $N$ varies and $K=4$.}
	\label{fig3}
\end{figure}
Fig. \ref{fig3} illustrates the relationship between $M$ and the downlink sum rate, where $N$ varies. Obviously, the achievable downlink sum rate grows when more legitimate APs are deployed. However, the presence of even a small number of adversarial APs compared to the number of legitimate APs can result in significant performance degradation. One can intuitively view the ultimate communications rate as the outcome of a game between legitimate nodes and illegitimate nodes. As shown in Fig. \ref{fig2}, in terms of reducing the achievable rate, assaulting the downlink data phase is more effective than the downlink PSA. It is worth noting that the rate loss resultant from attacking downlink data transmission phase can be decreased by raising $\rho_{\rm d}$. By contrast, raising $\rho_{\rm d}$ might be less effective in the presence of downlink PSAs because the downlink channel estimation get contaminated. 

\begin{figure}
	\centering
	\includegraphics[width=85mm]{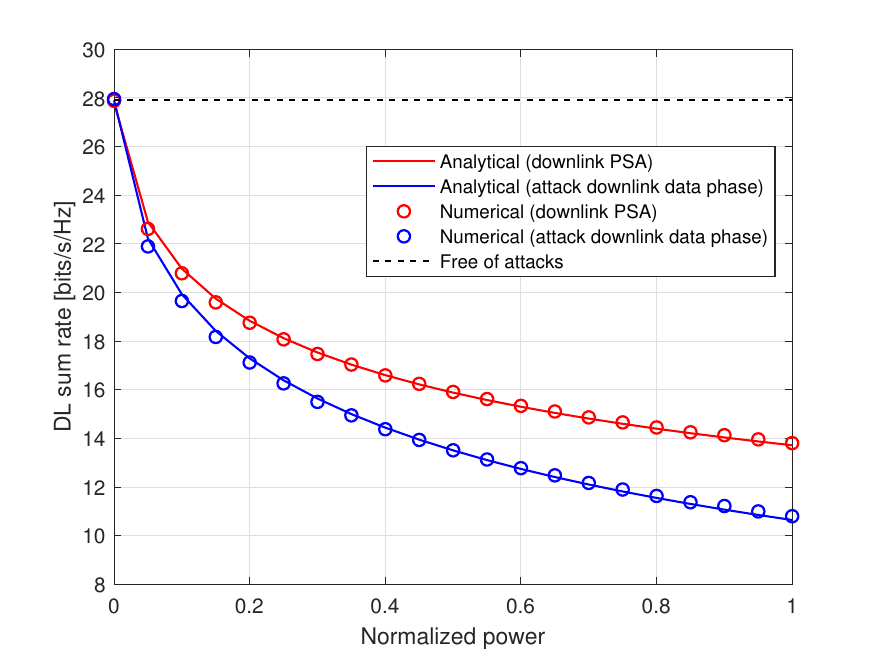}
	\caption{Downlink achievable sum rate versus the transmit power of adversarial APs, where $M=128$, $N=32$, and $K=4$.}
	\label{fig4}
\end{figure}

Moreover, Fig. \ref{fig4} demonstrates how the achievable downlink rate is impacted by the transmit power of adversarial APs when sending downlink beamformed pilots, with $M=128$, $N=32$, and $K=4$. The normalized power is calculated as $\mu_{\rm dp}/\rho_{\rm dp}$. As expected, the sum rate decreases as the transmit power of adversarial APs increases, but this also increases their likelihood of being detected. In order to strike a balance, a reasonable approach is to deploy more adversarial APs with lower transmit power. As an alternative, one can optimize $\zeta_{nk}$ to minimize the maximum $R_k$ using the max-min criterion.

\begin{figure}
	\centering
	\includegraphics[width=85mm]{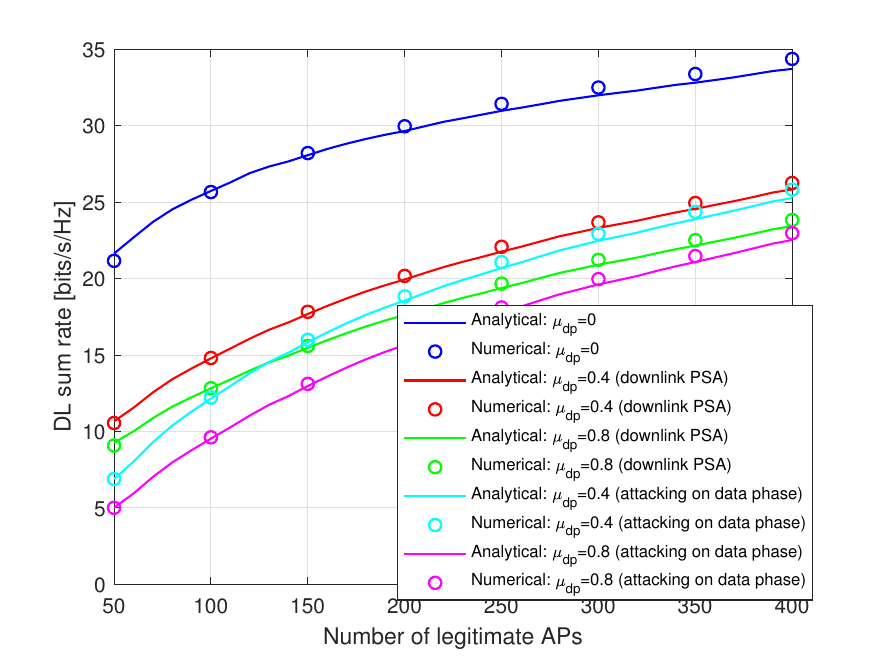}
	\caption{Downlink achievable sum rate versus the number of legitimate APs, where $\mu_{\rm dp}$ varies, $N=64$ and $K=4$.}
	\label{fig5}
\end{figure}

At last, Fig. \ref{fig5} illustrates the relationship between $R_{\rm sum}$ and $M$, where $\mu_{\rm dp}$ varies, with $N=32$ and $K=4$. Once again, the close agreement between analytical and numerical results confirms that \eqref{fr} and \eqref{fr1} are powerful tools for analyzing the performance of cell-free mMIMO systems in the presence of malicious attacks. As $\mu_{\rm dp}$ increases, the sum rate decreases accordingly. It is worth noting that adversarial APs have several options to launch effective attacks, such as modifying $M$, $\mu_{\rm dp}$, or $\zeta_{nk}$, making designing effective countermeasures quite challenging. As before, under the same setting, attacking the downlink data phase can cause more damage to the achievable rate compared to the downlink PSA. The intuitive understanding is that attacking the downlink data transmission is a direct approach compared with the downlink PSA. 

\subsection{Power Allocation of Adversarial APs}
In this subsection, we investigate the performance of power allocation from the perspective of adversarial APs. We focus on two attack strategies in particular, and for each of them, we take into account power allocation using equal and min-max optimization.

\begin{figure}
	\centering
	\includegraphics[width=85mm]{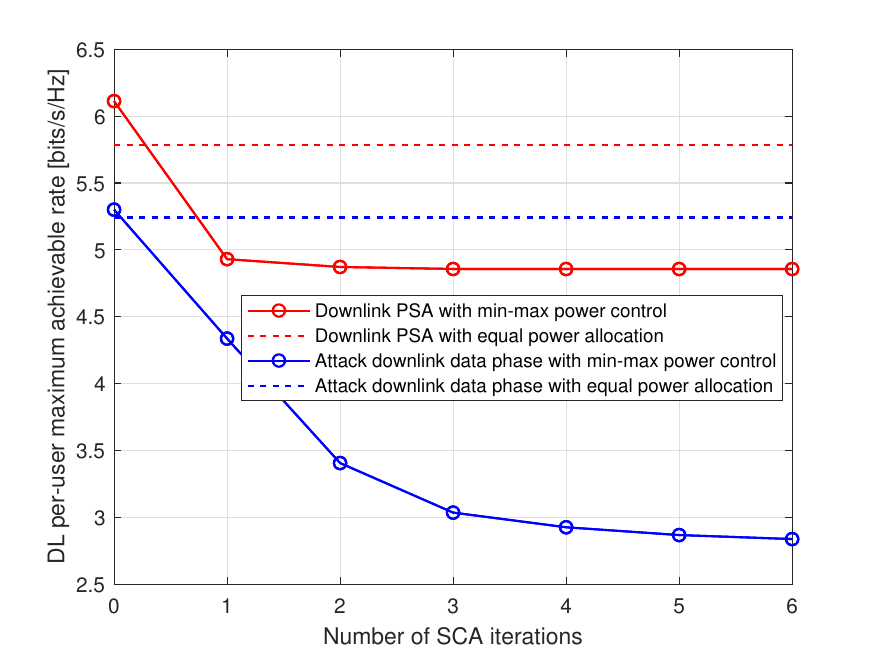}
	\caption{Relationship between $R_k^{\max}$ and the number of SCA iterations, where $M=128$, $N=32$, and $K=4$.}
	\label{fig6}
\end{figure}

Fig. \ref{fig6} depicts the maximum of the per-user achievable downlink rate $R_k^{\max}$ versus the number of SCA iterations. First, our proposed optimization approach proves to be much more effective than the equal power allocation scheme. Second, $R_k^{\max}$ decreases as the number of iterations increases, eventually stabilizing after reaching certain thresholds. In addition, different system parameters and channel fading models will result in different rate performance and convergence behavior. In particular, $R_k^{\max}$ converges to stable values after a single iteration in the case of the downlink PSA. However, in the case of attacking the downlink data transmission phase, it takes five iterations. Furthermore, the performance gap between these two attack strategies widens with an increasing number of iterations. For example, the rate gap is approximately 0.6 bits/s/Hz for one iteration, and rises to 2 bits/s/Hz for six iterations. We note that it would be important to research the scenario in which adversarial APs divide their transmit power between the training and data phases of the downlink in order to reduce the achievable rate of legitimate communications.

\begin{figure}
	\centering
	\includegraphics[width=85mm]{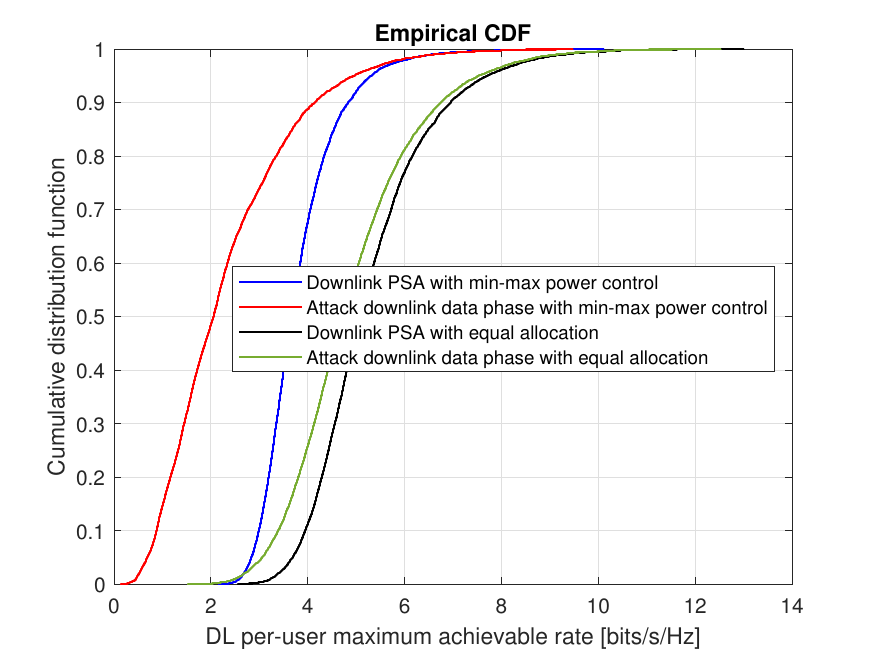}
	\caption{Cumulative distribution function of $R_k^{\max}$, where $M=128$, $N=64$, and $K=4$.}
	\label{fig7}
\end{figure}

Fig. \ref{fig7} plots the cumulative distribution function (CDF) of $R_k^{\max}$, where the scenarios under equal power allocation and min-max power control are included. Clearly, the min-max power control is the most effective approach for reducing the maximum per-user achievable rate, while the downlink PSA with equal power allocation is the least effective one. It is observed that the maximum per-user achievable rate with min-max power control is virtually below 8 bits/s/Hz, whereas that with equal power allocation is almost lower than 10 bits/s/Hz. The results in this figure can be viewed as a lower bound for $R_k^{\max}$, because solving $\mathbf{OP1.3}$ or $\mathbf{OP2.3}$ requires the knowledge of large-scale fading coefficients $\beta_{mk}$ and power allocation factors $\eta_{mk}$ of legitimate communications. However, acquiring these types of information would pose a challenge for adversarial APs.

\section{Conclusion}\label{conclusion}
The major finding of this paper is that the downlink training phase is just as vulnerable to PSAs as the uplink training phase in cell-free mMIMO systems. Specifically, we investigated the impact of downlink PSAs on the per-user achievable rate and derived its closed-form expression. It is evident that adversarial APs have flexible options to launch efficient attacks by modifying their numbers, transmit power, and power allocation factors. We conducted numerical simulations to validate our theoretical analysis, and both analytical and simulation results confirmed our findings.

If the large-scale coefficients and power allocation factors related to legitimate communications are known, we showed that adversarial APs can further reduce the per-user achievable rate by optimizing the downlink power allocation factors with the min-max power control criterion. We also investigated the situation where adversarial APs precoded random interference in the downlink data transmission phase in order to disrupt legitimate communications. Numerical results showed that although both schemes degrade the achievable downlink rate, attacking the downlink data phase is more detrimental than the downlink PSA. 

\appendices
\section{Derivation of Achievable Rate in \eqref{fr}}\label{app1}
According to \eqref{r_k2}, one must calculate $\mathbb{E}\small\{\left|\hat{\bar{a}}_{kk}\right|^2\small\}$ in order to determine the achievable per-user downlink rate. Given \eqref{channel-estimate-eve-1}, this task requires deriving $\mathbb{E}\small\{\small|{y}_{{{\rm dpa},k}}-\mathbb{E}\left\{{y}_{{{\rm dp},k}}\right\}\small|^2\small\}$, which is
\begin{equation}\label{appendix.1}
	\begin{aligned}		
		&~\mathbb{E}\left\{\left|{\bar{y}}_{{{\rm dp},k}}-\mathbb{E}\left\{{y}_{{{\rm dp},k}}\right\}\right|^2\right\}\\
		=&~\mathbb{E}\left\{{\bar{y}}_{{{\rm dp},k}}^{2} - 2{\bar{y}}_{{{\rm dp},k}}\mathbb{E}\left\{{y}_{{{\rm dp},k}}\right\} + \mathbb{E}\left\{{y}_{{{\rm dp},k}}\right\}^2\right\}\\
		=&~\mathbb{E}\left\{{\bar{y}}_{{{\rm dp},k}}^2\right\} - 2\mathbb{E}\left\{{\bar{y}}_{{{\rm dp},k}}\right\}\mathbb{E}\left\{{y}_{{{\rm dp},k}}\right\} + \mathbb{E}\left\{{y}_{{{\rm dp},k}}\right\}^2.
	\end{aligned}
\end{equation}
The expectations in \eqref{appendix.1} can be derived as
\begin{equation}
	\begin{aligned}
		\mathbb{E}\left\{{\bar{y}}_{{{\rm dp},k}}^2\right\} &= \tau_{\mathrm{d}} \rho_{\mathrm{dp}}\xi_k + \tau_{\mathrm{d}} \rho_{\mathrm{dp}}\varepsilon_k^2 +\\
		&+ \tau_{\mathrm{d}}\mu_{\mathrm{dp}}\sum\limits_{m=1}^{M}\eta_{mk}\beta_{mk}\gamma_{mk}\\
		&+ \tau_{\mathrm{d}}\mu_{\mathrm{dp}}\left(\sum\limits_{n=1}^{N}\sqrt{\zeta_{nk}}\kappa_{nk}\right)^2\\
		&+ 2 \tau_{\mathrm{d}} \varepsilon_k \sqrt{\rho_{\mathrm{dp}}\mu_{\mathrm{dp}}}\sum\limits_{n=1}^{N}\sqrt{\zeta_{nk}}\kappa_{nk}+1,\\
		\mathbb{E}\left\{{y}_{{{\rm dpa},k}}\right\}\mathbb{E}\left\{{y}_{{{\rm dp},k}}\right\} & = {\tau_{\mathrm{d}}\rho_{\mathrm{dp}}}\varepsilon_k^2\\
		&+\tau_{\mathrm{d}}\varepsilon_k\sqrt{\rho_{\mathrm{dp}}\mu_{\rm dp}}\sum\limits_{n=1}^N\sqrt{\zeta_{nk}}\kappa_{nk},\\
		\mathbb{E}\left\{{y}_{{{\rm dp},k}}\right\}^2 &= {\tau_{\mathrm{d}}\rho_{\mathrm{dp}}}\varepsilon_k^2.
	\end{aligned}
\end{equation}
Therefore, $\mathbb{E}\small\{\small|{\bar{y}}_{{{\rm dp},k}}-\mathbb{E}\left\{{y}_{{{\rm dp},k}}\right\}\small|^2\small\}$ can be obtained.

With $\mathbb{E}\small\{\small|{\bar{y}}_{{{\rm dp},k}}-\mathbb{E}\left\{{y}_{{{\rm dp},k}}\right\}\small|^2\small\}$, the expectation of $\left|\hat{\bar{a}}_{kk}\right|^2$ is then expanded into
\begin{equation}
	\begin{aligned}		
		\mathbb{E}\left\{\left|\hat{\bar{a}}_{kk}\right|^2\right\} &=  2A_1\mathbb{E}\left\{a_{kk}\right\}\left(\mathbb{E}\left\{{\bar{y}}_{{{\rm dp},k}}\right\}-\mathbb{E}\left\{{y}_{{{\rm dp},k}}\right\}\right)\\
		&+A_1^2\mathbb{E}\left\{\left|{\bar{y}}_{{{\rm dp},k}}-\mathbb{E}\left\{{y}_{{{\rm dp},k}}\right\}\right|^2\right\}+\mathbb{E}\left\{\left|{a}_{kk}\right|^2\right\}
	\end{aligned}
\end{equation}
where
\begin{equation*}
	A_1 = \frac{{\rm Cov}\left\{a_{kk},{y}_{{{\rm dp},k}}\right\}}{{\rm Cov}\left\{{y}_{{{\rm dp},k}},{y}_{{{\rm dp},k}}\right\}}.
\end{equation*}
According to \eqref{appendix.1} and with some algebraic manipulations, it then arrives at
\begin{equation}\label{app.2}
	\begin{aligned}
		&\mathbb{E}\left\{\left|\hat{\bar{a}}_{kk}\right|^2\right\} = \frac{2\varepsilon_k \upsilon_k \xi_k \tau_{\mathrm{dp}}\sqrt{\rho_{\mathrm{dp}} \mu_{\rm dp}}}{\tau_{\rm dp}\rho_{\rm dp}\xi_k + 1} + \varepsilon_k^2 \\
		&+ \frac{\tau_{\mathrm{dp}} \rho_{\mathrm{dp}} \xi^2}{\left(\tau_{\mathrm{dp}}\rho_{\mathrm{dp}}\xi_k +1\right)^2}\left(\tau_{\mathrm{dp}} \rho_{\mathrm{dp}} \xi_k + \tau_{\rm dp} \mu_{\rm dp} \left(\upsilon_k^2 + \xi_k\right)+1\right),
	\end{aligned}
\end{equation}
where $\upsilon_k = \sum\nolimits_{n=1}^{N}\sqrt{\zeta_{nk}}\kappa_{nk}$.

According to the definition of channel estimation error $\tilde{\bar{a}}_{kk}$, $\mathbb{E}\small\{\left|\tilde{a}_{kk}\right|^2\small\}$ can be expressed as 
\begin{equation}
	\begin{aligned}
		\mathbb{E}\left\{\left|\tilde{\bar{a}}_{kk}\right|^2\right\} &= A_1^2\mathbb{E}\left\{\left|{\bar{y}}_{{{\rm dp},k}}-\mathbb{E}\left\{{y}_{{{\rm dp},k}}\right\}\right|^2\right\} + \mathbb{E}\left\{\left|a_{kk}\right|^2\right\}\\
		& + 2A_1\mathbb{E}\left\{a_{kk}\right\}\left(\sqrt{\tau_{\mathrm{d}}\mu_{\rm dp}}\upsilon_k\right) -\mathbb{E}\left\{a_{kk}\right\}^2\\
		& - 2A_1\mathbb{E}\left\{a_{kk}\left({\bar{y}}_{{{\rm dp},k}}-\mathbb{E}\left\{{y}_{{{\rm dp},k}}\right\}\right)\right\}.
	\end{aligned}
\end{equation}
After tedious mathematical manipulations, $\mathbb{E}\small\{\left|\tilde{\bar{a}}_{kk}\right|^2\small\}$ can be obtained as
\begin{equation}\label{appendix.3}
	\begin{aligned}
		\mathbb{E}\small\{\left|\tilde{\bar{a}}_{kk}\right|^2\small\} &= - \frac{2\tau_{\rm dp}\rho_{\mathrm{dp}}\xi_k^2}{\tau_{\mathrm{dp}}\rho_{\mathrm{dp}}\xi_k + 1} + \frac{\tau_{\mathrm{dp}}^2 \rho_{\mathrm{dp}}^2 \xi_k^3 + \tau_{\mathrm{dp}}\rho_{\mathrm{dp}}\xi_k^2}{\left(\tau_{\mathrm{dp}}\rho_{\mathrm{dp}}\xi + 1\right)^2}\\
		& + \frac{\tau_{\mathrm{dp}}^2 \rho_{\mathrm{dp}} \mu_{\rm dp} \xi_k^2}{\left(\tau_{\mathrm{dp}}\rho_{\mathrm{dp}}\xi_k+1\right)^2}\left(\upsilon_k^2 + \xi_k\right) + \xi_k.
	\end{aligned}
\end{equation}
With $\mathbb{E}\left\{\small|a_{kk'}\small|^2\right\} = \sum\nolimits_{m=1}^M \eta_{m{k'}} \beta_{mk'} \gamma_{m{k'}}$, combining results in \eqref{app.2} and \eqref{appendix.3} leads to the expression of SINR $\lambda_{k}$, and then \eqref{fr} is obtained.

\section{Derivation of Achievable Rate in \eqref{rate_data_1}}\label{app2}
First, we have the following expectation
\begin{equation}
	\begin{aligned}
		\mathbb{E}\small\{\left|\hat{a}_{kk}\right|^2\small\} &= \left|\mathbb{E}\small\{a_{kk}\small\}\right|^2 + A_1^2\mathrm{Var}\left\{y_{\mathrm{dp},k}\right\}\\
		& = \left(\sum\limits_{m=1}^{M}\sqrt{\eta_{mk}}\gamma_{mk}\right)^2 + \kappa_{{k}}		
	\end{aligned}
\end{equation}
where $\kappa_{k} = \frac{\tau_{\mathrm{dp}}\rho_{\rm dp}\varsigma_{kk}^2}{1 + \tau_{\mathrm{dp}}\rho_{\mathrm{dp}}\varsigma_{kk}}$.

Next, the expectation of $\left|\tilde{a}_{kk}\right|^2$ can be shown as
\begin{equation}
	\mathbb{E}\left\{\left|\tilde{a}_{kk}\right|^2\right\} = \mathbb{E}\left\{\left|{a}_{kk} - \hat{a}_{kk}\right|^2\right\} = \varsigma_{kk} - \kappa_{{k}}.
\end{equation}
Then, the expectation of $\left|{a}_{kk'}\right|^2$ is calculated by
\begin{equation}
	\begin{aligned}
		&\mathbb{E}\left\{\left|{a}_{kk'}\right|^2\right\}\\
		&= \sum_{m=1}^{M} \eta_{mk'}\mathbb{E}\left\{\left|g_{mk}\hat{g}^*_{mk'}\right|^2\right\}\\
		&+ \sum_{m=1}^{M}\sum_{n \ne m}^{M} \sqrt{\eta_{mk'}\eta_{nk'}}\mathbb{E}\left\{g_{mk}g_{nk}^*\hat{g}^*_{mk'}\hat{g}_{nk'}\right\}\\
		& = \sum_{m=1}^M\eta_{mk'}\beta_{mk}\gamma_{mk'}.
	\end{aligned}
\end{equation}
And finally, the mean of $\left|b_{kk}\right|^2$ can be computed by
\begin{equation}
	\mathbb{E}\left\{\left|b_{kk}\right|^2\right\} = \left(\sum_{n=1}^{N}\sqrt{\zeta_{nk}}\kappa_{nk}\right)^2 + \sum_{k'=1}^K\sum_{n=1}^{N}\zeta_{nk'}\theta_{nk}\kappa_{nk'}.
\end{equation}
With all the results above, the achievable rate in \eqref{rate_data_1} is then obtainable.

\bibliographystyle{IEEEtran}
\bibliography{reference.bib}

\vspace{12pt}

\end{document}